\documentclass[%
reprint,
superscriptaddress,
bibnotes,
amsmath,amssymb,
aps,
prb,
]{revtex4-2}

\usepackage{graphicx}
\usepackage{dcolumn}
\usepackage{bm}
\usepackage{multirow}
\usepackage{booktabs} 
\usepackage[normalem]{ulem}
\usepackage{hyperref}

\usepackage[usenames,dvipsnames]{color}

\newcommand{\be}{\begin{equation}}
\newcommand{\ee}{\end{equation}}
\newcommand{\bea}{\begin{eqnarray}}
\newcommand{\eea}{\end{eqnarray}}

\begin{document}

\preprint{APS/123-QED}

\title{Excitonic response in transition metal dichalcogenide heterostructures from first-principles: Impact of stacking, twisting,  and interlayer distance}

\author{R. Reho}
 \email{r.reho@uu.nl}
\affiliation{
Debye Institute for Nanomaterial Science, Utrecht University, Princetonplein 5, 3584 CC Utrecht, The Netherlands and ETSF
}%

\author{A. R. Botello-Méndez}
\affiliation{
Debye Institute for Nanomaterial Science, Utrecht University, Princetonplein 1, 3584 CC Utrecht, The Netherlands and ETSF
}

\author{D. Sangalli}
\affiliation{CNR-ISM, Division of Ultrafast Processes in Materials (FLASHit), Area della Ricerca di Roma Tor Vergata, 100 Via del Fosso del Cavaliere, Rome, Italy and ETSF}

\author{M. J. Verstraete}
\affiliation{
Debye Institute for Nanomaterial Science, Utrecht University, Princetonplein 1, 3584 CC Utrecht, The Netherlands and ETSF
}
\affiliation{nanomat/Q-mat, Université de Liège, B-4000 Sart Tilman, Liège, Belgium}

\author{Zeila Zanolli}

\affiliation{
Debye Institute for Nanomaterial Science, Utrecht University, Princetonplein 5, 3584 CC Utrecht, The Netherlands and ETSF
}
\date{\today}

\begin{abstract}
Van der Waals heterostructures of two-dimensional transition metal dichalcogenides provide a unique platform to engineer optoelectronic devices tuning their optical properties via stacking, twisting, or straining.
Using \textit{ab initio} many-body perturbation theory, we predict the electronic and optical (absorption and photoluminescence spectra) properties
of MoS$_2$/WS$_2$ and MoSe$_2$/WSe$_2$ heterobilayers with different stacking and twisting.
We analyze the valley splitting and optical transitions, and we explain the enhancement or quenching of the inter- and intra-layer exciton states.
We fully include transitions within the entire Brillouin Zone,  contrary to predictions based on continuum models which only consider energies near the $K$ point. As a result, we predict an interlayer exciton with significant electron density in both layers and a mixed intralayer exciton distributed over both MoSe$_2$ and WSe$_2$ in a twisted Se-based heterostructure.
We propose that it should be possible to produce an inverted order of the excitonic states in some MoSe$_2$/WSe$_2$ heterostructures, where the energy of the intralayer WSe$_2$ exciton is lower than that in MoSe$_2$.
We predict the variability across different stacking of the exciton peak positions ($\sim$100 meV) 
and the exciton radiative lifetimes, from pico- to nano-seconds, and even micro-seconds in twisted bilayers.
The control of exciton energies and lifetimes paves the way towards applications in quantum information technologies and optical sensing.
\end{abstract}
\maketitle


\section{Introduction}\label{sec:intro}

\begin{figure}[t]
    \includegraphics[width=\columnwidth]{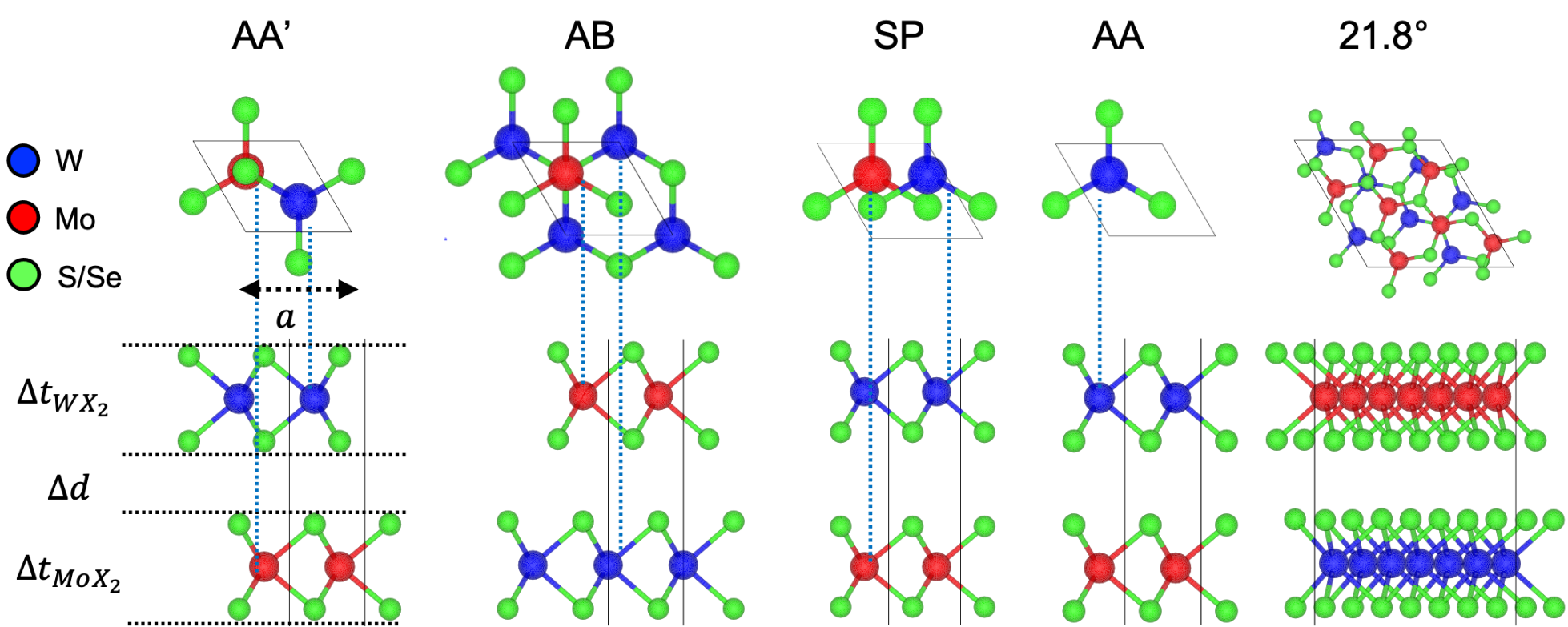}
    \caption{\label{fig:geometricalmodels}Top and side views of vdW HS with lattice parameter $a$,  van der Waals gap $\Delta d$, and monolayer thickness $\Delta t_{\rm{YX_2}}$ 
    with Y = W, Mo and X = S, Se.
    Relaxed structural parameters are reported in Table~\ref{tab:infogeometryHS}.}
\end{figure}

Transition metal dichalcogenides (TMDs) can host strong light-matter interaction and enhanced excitonic effects, making them ideal for next-generation optical devices~\cite{wilson2021excitons}.
TMD monolayers (MLs) can be further stacked into van der Waals heterostructures (vdW HS)~\cite{novoselov2005two, splendiani2010emerging, lee2012synthesis} with added functionalities.
Their electronic and optical properties strongly depend on layer thickness ($\Delta t$), interlayer distance ($\Delta d$), stacking and twist angle~\cite{sohier2023impact, tran2019evidence, rakib2022moire}.
Achieving experimental control on these features is complex~\cite{ma2023small, ribeiro2018twistable}:
encapsulation affects thickness and interlayer distance, and the dielectric environment influences transport and other properties. 
Therefore, theoretical models are essential for isolating effects and guiding experimental design.
In addition, for small twist angles, 
TMDs HS display regions which are misaligned and other with local atomic registry (AA$^\prime$, AA, AB and
SP, Fig.~\ref{fig:geometricalmodels}). In general, the changes of these degrees of freedom occur simultaneously. This makes it experimentally difficult to disentangle them.

Heterostructures of TMDs with different transition metal atoms 
typically show a type II band alignment, which leads to the formation of interlayer excitons (IL),
in addition to intralayer (IN) ones~\cite{wang2018colloquium}. 
Understanding how IN and IL exciton energies and intensities are affected by the above degrees of freedom is essential to control and exploit 
them for quantum information technologies~\cite{ji2017robust, trovatello2020ultrafast, tsai2022antisite}. 
The energetics, intensity and dynamics of bright IN excitons can be probed via photoluminescence (PL) or time-resolved photoluminescence (TRPL)~\cite{wang2018colloquium,jiang2021interlayer,trovatello2020ultrafast}.
Establishing a correspondence between observed peaks and individual monolayers
is usually done by analogy with exciton energies in the freestanding MLs. 
However, this approach has limited predictive power as it neglects the vdW interaction between layers and the resulting electrostatic environment (screening).
\textit{Ab initio} many-body perturbation theory (Ai-MBPT) formalism, instead, allows for unambiguous identification of the spatial and spectral origin of each excitonic transition. 
The first \textit{ab-initio} calculations of PL and absorption by IL and IN excitons in a TMD vdW HS were done by Torun et al~\cite{torun2018interlayer}, but with a focus on the AA$^\prime$ stacking only.
The same year, Gillen et al.~\cite{gillen2018interlayer} computed absorption spectra of Se-based AA$^\prime$, AA, AB HS using the Bethe-Salpeter Equation (BSE).

In this work, we systematically analyze the variation of the electronic and optical properties of 
MoX$_2$/WX$_2$ (with X = S/Se)
vdW HS with interlayer distance, stacking and twisting, using Ai-MBPT~\cite{sangalli2019many, marini2009yambo}
We have updated the PL implementation in the \textsc{Yambo} code, following the 
general formalism of Melo and Marini~\cite{de2016unified} 
and incorporated the approaches and formalism developed in Refs.~\cite{lechifflart2023first, paleari2019exciton}.
These theories provide a framework for a full \textit{ab-initio} description of the coupled out-of-equilibrium dynamics of photons, phonons, and electrons at different level of approximation.
The dynamics of the carriers following light excitation can  either be described in real-time or as a noncoherent thermalized exciton population.
We implemented the computation of the lesser component of the electron-hole Green's function, $L$, 
which gives access to PL  and transient absorption~\cite{marini2009yambo,de2016unified}.
%
We find that,
under sufficient compression along the stacking direction, the excitonic ground state originates from an emerging transition at $\Gamma$. This is due to enhanced orbital overlap in the vdW gap.
In-plane compressive (respectively, tensile) strain results in an increase (respectively, decrease) of the first and second bright exciton energies.
We predict that absorption and PL spectra can be unequivocally attributed to specific stacks/twists, as the corresponding energy ordering of exciton energies are of the order of $\sim$100 meV. 
The absorption and PL spectra of twisted MoSe$_2$/WSe$_2$ heterostructures is particularly interesting as it hosts new emergent features: 
flat bands, an IL exciton with significant electron density contribution in both layers, two degenerate IN excitons localized on each of the ML, suppression of IN WSe$_2$ exciton, and reversed excitonic order.  
The MoSe$_2$/WSe$_2$ twisted heterostructure is promising for applications in photovoltaics, as the computed exciton lifetimes reach 0.4 $\mu$s, namely several order of magnitudes larger than typical values~\cite{palummo2015exciton, chen2018theory}. 

\section{Theoretical Methods}\label{sec:theory}
The ground state properties 
are computed by solving the Kohn-Sham (KS) equations within the density functional theory (DFT) formalism using the Quantum Espresso (QE) code~\cite{giannozzi2009quantum, giannozzi2017advanced}.
All structural relaxations were performed including spin-orbit Coupling (SOC),  semiempirical van der Waals correction D2~\cite{barone2009role}, and Se semicore states in the pseudopotential.
The inclusion of Se semicore states is essential to predict the direct band-gap of MLs. 
Usually, the 3s and 3p Se states are assumed to be frozen in the core because of the large energy separation from 3d states. However, the spatial overlap of the corresponding wavefunctions is nonnegligible. Freezing the semicore states results in an unphysical shift of the extrema of the conduction and valence bands, and indirect band-gap for both MoSe2 and WSe2 MLs phases (Appendix~\ref{app:pseudos}).

To obtain quantitative predictions of the electronic structure, we computed
corrections to DFT energies using the GW method~\cite{hedin1965new} in the plasmon-pole approximation for all 
$\textbf{k}$-points in the Brillouin Zone (BZ)~\cite{farid1988gw}.
We employed the stochastic integration and the 2D cut-off of the Coulomb potential~\cite{guandalini2023efficient} to compute the screening $W$.
Neutral excitations are computed from the BSE for electron-hole interactions in the Ai-MBPT formalism~\cite{sangalli2019many, marini2009yambo}.
To obtain the exciton energies and wavefunctions, we solve the BSE in the Tamm-Dancoff approximation including local field effects~\cite{onida2002electronic}. 
The BSE can be recast into an eigenvalue equation 
\be \label{eq:BSE}
    \left(\varepsilon_{c\mathbf{k}}^{\mathrm{GW}}
    -\varepsilon_{\mathrm{v}\mathbf{k}}^{\mathrm{GW}}\right)
    A_{\mathrm{vc}\mathbf{k}}^\lambda+\sum_{\mathbf{k}^{\prime}
    c^{\prime} \mathrm{v}^{\prime}}
    K_{\substack{\mathrm{vc\mathbf{k}} \\ \mathrm{v}^{\prime} \mathrm{c}^{\prime} \mathbf{k}^{\prime}}}^{\mathrm{eh}}
    A_{\mathrm{v}^{\prime} \mathrm{c}^{\prime} \mathbf{k}^{\prime}}^\lambda
    =E_\lambda A_{\mathrm{v c} \mathbf{k}}^\lambda
\ee
where $\varepsilon^{GW}
_{\mathrm{c}\mathbf{k}/\mathrm{v}\mathbf{k}}$ are quasiparticle band energies, 
 $A^{\lambda}_{\mathrm{v c}\mathbf{k}}$ are the BSE coefficients and $E_{\lambda}$ 
the energy of exciton $\lambda$. The optical absorption is computed from the imaginary part
of the polarizability function $\alpha_{\mathrm{2D}}(\omega)$, given by

\be\label{eq:polarizability}
\alpha_{\mathrm{2D}}(\omega)=\frac{1}{A_{\mathrm{uc}}} \sum_\lambda \frac{\left|\mu_\lambda\right|^2}{\omega-E_\lambda+i \eta}
\ee
where $A_{\mathrm{uc}}$ is the area of the unit cell.~\cite{marini2009yambo,sangalli2017optical}.
The exciton transition dipole, $\mu_{\lambda}$, is defined in terms of the electronic dipoles as 
$\mu_{\lambda} = \sum_{\mathrm{cv}\mathbf{k}} A^{\lambda}_{\mathrm{cv}\mathbf{k}} d_{\mathrm{cv}\mathbf{k}}$, with $d_{\mathrm{cv}\mathbf{k}} = \langle \mathrm{v}\mathbf{k}|e\mathbf{r}|\mathrm{c}\mathbf{k}\rangle$, $e$ the electron charge and $\mathbf{r}$ the one-body position operator.

%
The PL spectra are computed as in Ref.~\cite{de2016unified}, analogously to 
Refs~\cite{libbi2022phonon, torun2018interlayer, paleari2019exciton, lechifflart2023first}. 
Here, we neglect the exciton-phonon coupling. 
Note that, the interaction between excitons and phonons helps to overcome the momentum mismatch between the electron and hole, and leads to changes in the spectra . State of the art calculations on the exciton-phonon coupling in Photoluminescence~\cite{paleari2019exciton,lechifflart2023first}  are beyond the scope of the paper. They will be exceedingly heavy, as they require to compute the excitonic energies and wave-functions at finite momentum, as well as the electron-phonon matrix elements.
In this approximation, the intensity of the emission spectra $I^{\mathrm{PL}}$ is
\be\label{eq:PL}
I^{\mathrm{PL}}(\omega)=
\frac{\omega^3 n_r(\omega)}{\pi^2 \hbar c^3}
\sum_\lambda \mathfrak{Im}\left[ \frac{\left|\mu_\lambda\right|^2}{\omega-E_\lambda+i \eta} \right]
e^{-\frac{E_\lambda-E_{\text {min}}}{k_BT_{\mathrm{exc}}}}
\ee
where,
$E_{\mathrm{min}}$  the minimum energy of the exciton dispersion, $n_r(\omega)$$\approx 1$ the refractive index for a free-standing 2D material, 
$i\eta$ an infinitesimally small imaginary number, and $k_B$ the Boltzmann constant.
We assume a low-density thermalized initial population of excitons.
This can be modelled by a Bose distribution function, which parametrically depends on the exciton temperature $T_{\mathrm{exc}}$~\cite{hong2014ultrafast, heo2015interlayer, chen2016ultrafast, torun2018interlayer}.
The excited state dynamics is described by the exciton radiative lifetimes $\tau_{\lambda}(0)$
in the optical limit ($\mathbf{q}\rightarrow 0$) ~\cite{palummo2015exciton,chen2018theory,li2023strain}:
\begin{equation}\label{eq:tau_exc}
    \tau_{\mathrm{\lambda}}(0)=\frac{\hbar^2 c A_{\mathrm{uc}}}{4 \pi e^2 E_{\mathrm{\lambda}}\mu_{\mathrm{\lambda}}^2}{} .
\end{equation}

To visualize exciton wavefunctions ($\Psi$) or identify IN or IL cases, one often chooses a ``chemically reasonable'' position for the hole (e.g. on the electrophilic chalcogen) and examines the electron part of $\Psi$. In the HS in particular, this choice is delicate and potentially arbitrary, e.g. which layer's chalcogen should we choose? 
To analyze electron and hole localization within the heterostructure we define the probability $P(\mathrm{IN})$ that a transition is intralayer using the integrals
\be\label{eq:INcoeff}
P^{\lambda, \lessgtr}(\mathrm{IN}) = \int_{z_{\mathrm{e}} \; \mathrm{and} \; z_{\mathrm{h}} \lessgtr z_0}
\left|\Psi^\lambda(\mathbf{r}_{\mathrm{e}}, \mathbf{r}_{\mathrm{h}}) \right|^2 d \mathbf{r}_{\mathrm{e}} \mathrm{d} \mathbf{r}_\mathrm{h} 
\ee
where $\Psi^\lambda = \sum_{\mathrm{vc}\mathbf{k}} A^{\lambda}_{\mathrm{vc}\mathbf{k}} |\mathrm{v}\mathbf{k}\rangle|\mathrm{c}\mathbf{k}\rangle$ is the exciton wavefunction, and  $\mathbf{r}_\mathrm{e}$, $\mathbf{r}_\mathrm{h}$ denote electron and hole coordinates, respectively, with the $z$ coordinate normal to the heterostructure.
The threshold $z_0$ is set halfway 
between the two layers.
The $<$ and $>$ in Eq.~(\ref{eq:INcoeff}) denote that integration is done for $z_h$ smaller or greater than the threshold $z_0$.
It follows that $P^\lessgtr(\mathrm{IL}) = 1-P^\lessgtr(\mathrm{IN})$ and $P(\mathrm{IL}) = P^<(\mathrm{IL})+P^>(\mathrm{IL})$ give the interlayer transition probabilities. Their values are reported in Tables~\ref{tab:mos2ws2PIL} and~\ref{tab:mose2wse2PIL}. 
The $P(\mathrm{IN})$ defined above is unequivocal in this sense and can distinguish IN and IL for each layer and in a quantitative fashion.
The implementation of Eq.~(\ref{eq:tau_exc}) and other post-processing to reproduce the main results of this work are available in the author's forked repository of the \textsc{Yambopy} Python library \footnote{https://github.com/rreho/yambopy}.

\section{Computational details}\label{sec:compdetails}
All calculations have been performed with fully relativistic pseudopotentials from PseudoDojo including the s, p and d 
semicore shells~\cite{vanSettenPseudoDojoTraining2018}. The ground state properties have been computed with Quantum Espresso
\cite{giannozzi2009quantum} relaxing the structures with the Broyden (BFGS~\cite{head1985broyden}) algorithm and the inclusion of spin orbit coupling and vdW corrections~\cite{barone2009role,grimme2010consistent}, 
a plane wave kinetic energy cutoff of 120 Ry, and the PBE~\cite{PerdewGeneralizedGradient1996} exchange correlation functional.
Atomic forces and pressure were converged to better than  $10^{-7}$ Ry/a.u. and $0.5$ kbar, respectively.
With this choice, the total energies are converged to within less than 1 meV per unit cell.

The GW correction and excitonic properties have been computed on a 30x30x1 grid with 12 \AA ~of vacuum for the MLs and 20 \AA ~of vacuum for the HS. We employed the stochastic integration and the 2D cut-off of the Coulomb potential described in Ref.~\cite{guandalini2023efficient} to reduce the needed size of the reciprocal space grid.
We included 500 unoccupied bands for the MLs, 900 for the S-based HS, 1200 for the Se-based aligned HS and 1600 for the twisted MoSe$_2$/WSe$_2$ to compute the screening and the BSE kernel.
The values of the kinetic energy cut-off for the response functions and 
the exchange energy were converged to ensure precision better than $\sim$50 meV.
Unless otherwise stated, we employed a $T_{\mathrm{exc}}=\mathrm{400}$ K for the exciton population, which allows us to stay within the low exciton density approximation (\ref{sec:theory}) and clearly resolve the emission peaks in PL, which inevitably depend on the number of available exciton states.

\section{Results}\label{sec:results}

\subsection{Ground state properties}\label{subsec:geometry}
All MLs and HS were fully relaxed (atoms and cell) with stresses and forces below $0.5$ kbar and $10^{-5}$ Ry/a.u..
The lattice parameters (3.191 \AA~and 3.327 \AA~ for S- and Se-based HS, respectively), remained unchanged in the other stackings.
The most stable configuration is AA$^{\prime}$, followed by AB, SP and AA (Table~\ref{tab:tot-energies}). 
In the AA stacking the metal atoms belonging to different MLs are on top of each other, resulting in electrostatic repulsion, increased total energy and increased interlayer distance $\Delta d$. The layer thickness $\Delta t$ is not affected by the stacking (Table~\ref{tab:infogeometryHS}).
The twisted Se-based heterostructure has been constructed by rotating it 21.8$^\circ$ relative to the AA stacking configuration, with the rotation axis positioned above two hollow sites. This results in an atomic arrangement intermediate between AA and AB stacking, with 42 atoms in the moir\'e unit cell.
During relaxation of the twisted Se-based HS, instead, the layer spacing shrinks significantly, resulting in enhanced pressure along the stacking direction and higher total energy.
The twisted heterostructure 
could give indications on the physics emerging in misaligned configurations, 
as the periodically repeated unit cell is larger and the resulting electrostatic potential more complex than the un-twisted cases.

\begin{table}[t]
    \caption{\label{tab:tot-energies} Total energy differences per bilayer unit cell ($\Delta$E, meV) and interlayer distance ($\Delta d$, \AA) of TMD HS. 
    }
    \begin{tabular}{c|c|c|c|c|c|c}
        \hline
                 & & AA$^\prime$ & AB & SP & AA & Twisted \\
        \hline
        MoS$_2$/WS$_2$ & $\Delta$E & 0 & 3 & 21 & 75 & // \\
                       & $\Delta d$ & 3.040 & 3.026 & 3.523 & 3.633 & // \\
        MoSe$_2$/WSe$_2$ & $\Delta$E & 0 & 4 & 26  & 105 & 980 \\
                         & $\Delta d$ & 3.137 & 3.138 & 2.253  & 3.721 & 2.954 \\
        \hline  
    \end{tabular}
    \end{table}    

\begin{figure*}[t]
    \includegraphics[width=0.9\textwidth]{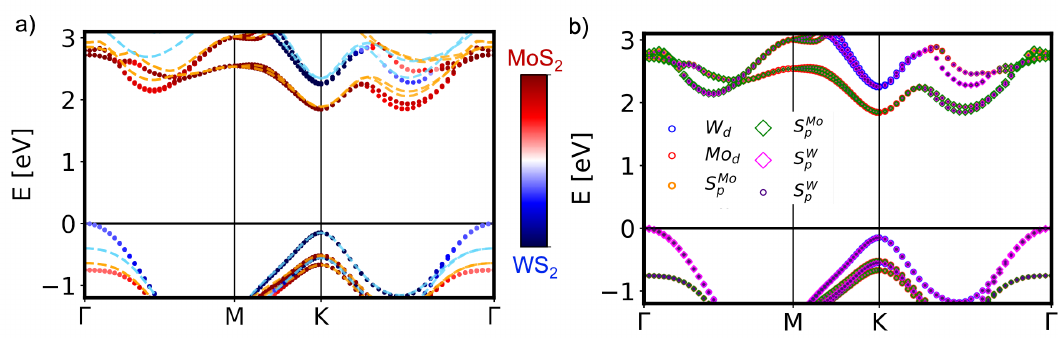}
    \caption{\label{fig:mos2ws2abbands} GW electronic band structure of the AB stacked MoS$_2$/WS$_2$ HS projected over the (a) MoS$_2$ (red) and WS$_2$ (blue) layers  or (b) individual atomic orbitals ${\rm{Y^{M}}}_b$ ($b$ orbital of atom Y in layer M). The orange and light blue dashed lines denote the  GW bands of the individual MLs. 
    The diamond-shaped symbols in panel (b) highlight the contribution from the orbitals of the chalcogen atoms close to the vdW gap.} 
\end{figure*}

\begin{figure}[b]
    \includegraphics[width=\columnwidth]{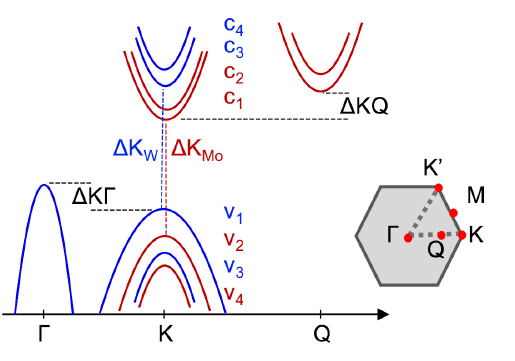}
    \caption{\label{fig:schematicgwbands} Schematic of a vdW HS band structure representing the layer projected bands (same color convention as in [Fig.~\ref{fig:mos2ws2abbands}(a)] and the valley splittings $\Delta KQ$ and $\Delta K\Gamma$.
    $\Delta K_{\mathrm{Mo}}$/$\Delta K_{\mathrm{W}}$ label the energy difference at $K$ between bands belonging to the same monolayer ($\mathrm{c}_1 - \mathrm{v}_2$ and $\mathrm{c}_3 - \mathrm{v}_1$,  respectively). Note that $\mathrm{c}_1 - \mathrm{c}_2$ and $\mathrm{c}_3 -\mathrm{c}_4$ are degenerate at $K$.} 
\end{figure}

\subsection{GW electronic band structure}\label{sec:bandstructure}
A quantitative description of the electronic band structure allows one to determine the spatial origin of optical transitions and the direct/indirect nature of each system. This is especially important in phonon assisted processes where multiple electron scatterings might occur from $K$ to $Q$ and vice versa.

In Fig.~\ref{fig:mos2ws2abbands} we analyze the GW band structure for the specific case of MoS$_2$/WS$_2$ in the AB stacking.
All GW band structures are reported in Appendix~\ref{app:GW} and Fig.~\ref{fig:moX2wX2-projbands}.
The bands are projected either on layers [Fig.~\ref{fig:mos2ws2abbands}(a)], as schematically represented in Fig.~\ref{fig:schematicgwbands}, or on specific atomic orbitals [Fig.~\ref{fig:mos2ws2abbands}(b)]. The orbital projection shows the effect of the MLs hybridization and screening in the HS. 
The HS band structure is more complex than a simple superposition of those of the constituent layers [orange and light blue dashed lines in Fig.~\ref{fig:mos2ws2abbands}(a)]: the valence and conduction band edges display a type II alignment at $K$, with a reduced fundamental gap at $K$ ($\Delta K$ = 2.058 eV) with respect to the MLs, leading to the formation of IL excitons. 
The GW gap of TMD MLs and HS ranges between $\sim$1.2 -- 2.5 eV and increases with the atomic number of the chalcogen atom, while it is less sensitive to 
the transition metal (Table~\ref{tab:infoHSsSOC}).

The electronic states close to $K$  either belong to WS$_2$ or MoS$_2$ layers [Figs.~\ref{fig:mos2ws2abbands}(a) and ~\ref{fig:schematicgwbands}]. Specifically, they belong to $p$ and $d$ orbitals of sulfur and metal atoms, respectively [Fig.~\ref{fig:mos2ws2abbands}(b)].
We denote $\Delta K_{M}$, M = (Mo, W) the layer-specific GW gaps (topmost valence to lowest conduction band belonging to W/Mo). These determine the continuum energy level of IN excitons and their binding energy (BE).
We predict that due to interlayer screening the $\Delta K_{\mathrm{Mo/W}}$ gap in a vdW HS is considerably lower than the monolayer one, except for the AA$^{\prime}$ MoS$_2$/WS$_2$ HS (Tables~\ref{tab:infoHSsSOC} and ~\ref{tab:infosplittingsMLs}).

All systems display a similar orbital projected band structure (Fig.~\ref{fig:moX2wX2-projbands}) since they share the same crystal symmetry. 
The differences are due to the details of the stacking:
The spatial overlap between out-of-plane $p$ and $d$ orbitals from each layer affects the valley splitting at high symmetry points of the conduction ($\Delta KQ = E_{Q}-E_{K}$) and valence ($\Delta K\Gamma = E_{K}-E_{\Gamma}$) bands and the SOC splitting at $K$ [Fig.~\ref{fig:mos2ws2abbands}(b)].
Going from the most stable (AA$^\prime$) to the less stable (AA) structure, 
the energy overlap between the second and third valence band (involving $p$ and $d$ orbitals from both layers) increases (Table~\ref{tab:infoHSsSOC}),
pushing upwards the valence band at the $K$ point and resulting in a direct gap for the AA structure ($\Delta K\Gamma > 0$). 
The Se-based HS exhibit direct gap, except for the twisted one.
The variations in the band structure influence the IL and IN exciton energies; however, these are not the only determinants of the absorption peaks (see Sec.~\ref{subsec:HSabspl}.b). The many-body Kernel, as described by Eq.~(\ref{eq:BSE}), introduces additional complexity through screening effects that differ with each stacking configuration.

\begin{figure*}[t]
    \includegraphics[width=0.9\textwidth]{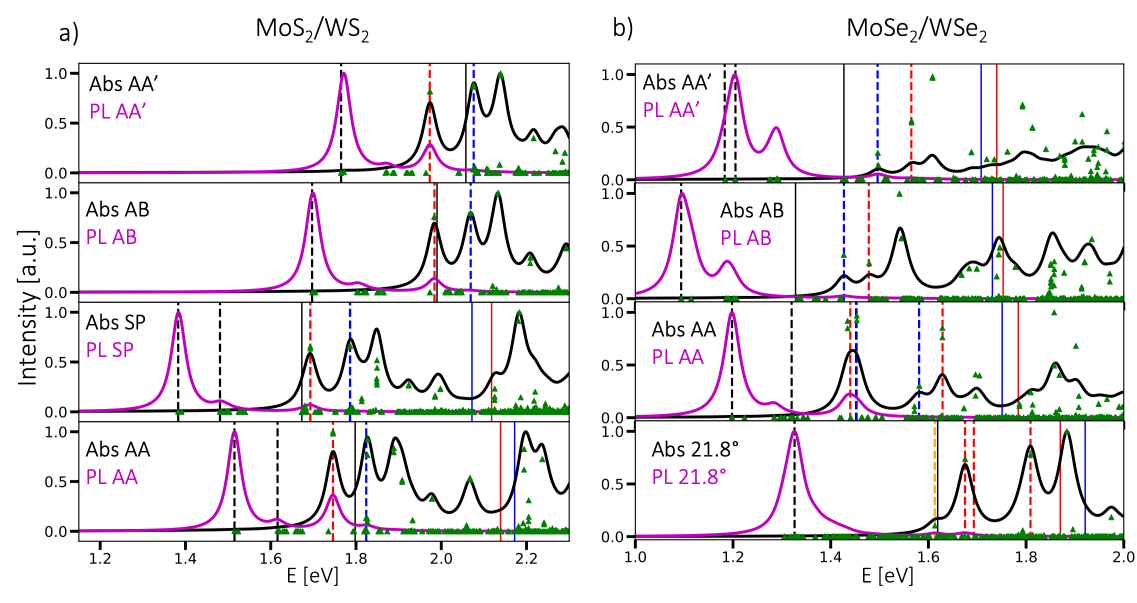}
    \caption{MoS$_2$/WS$_2$ and MoSe$_2$/WSe$_2$ absorption (black) and PL (magenta) spectra and oscillator strengths (green triangles). Spectra are arranged from top to bottom based on decreasing structural stability. Vertical dashed lines indicate the excitonic peak energies with their corresponding character: IL in black, IN-Mo in red, IN-W in blue, and IN(Mo+W) in orange.
    Vertical solid lines indicate the electronic gaps at $K$ of the full HS ($\Delta K$, black) and of the projected bands for the individual Mo ($\Delta K_{\mathrm{Mo}}$, red) and W layers ($\Delta K_{\mathrm{W}}$, blue).
    }    
    \label{fig:HSabspl4stackings}
\end{figure*}

The computed GW gap of S-based HS agree  with the 
literature within $\sim$10 meV, while
the Se-based gaps are red-shifted by about $\sim$100 -- 200 meV~\cite{splendiani2010emerging, mak2010atomically, chhowalla2013chemistry, zhang2014direct, zhao2013evolution, lan2018synthesis}. 
This discrepancy is within the typical accuracy of GW/BSE simulations and
can be attributed to specific simulation choices:  
relaxation with SOC,  full Se semicore pseudopotential, vdW correction, 
and use of the RIM-W 2D Coulomb potential
integration method~\cite{guandalini2023efficient}, which leads to increased screening 
with respect to previous simulations. 

\subsection{Optical properties of HS}\label{subsec:HSabspl}
In this section, we present the absorption and emission spectra of vdW HS. In particular, we discuss the IL excitons and the change in their energy onset upon 
varying the interlayer distance $\Delta d$. We distinguish between symmetry forbidden dark IL excitons and ``gray'' excitons, which have very small transition dipoles due to the reduced overlap between the hole and electron wavefunctions.
We number the excitonic peaks with IX$_N$ (where $N=1,2...$) . The binding energy (BE) is defined as $\text{BE} = \Delta K - \text{IX}_{N} $ for interlayer excitons and $\text{BE} = \Delta K_{\mathrm{Mo/W}} - \text{IX}_{\mathrm{N}} $  for intralayer excitons.

\paragraph{Spectra:}
Fig.~\ref{fig:HSabspl4stackings} shows the PL and absorption spectra [Eq.~(\ref{eq:polarizability})] for different HS stackings. 
For the stacking that are closer in total energy (AA$^\prime$, AB, AA SP), we predict a systematic energy ordering of the spectra; from $\text{AA}^\prime \rightarrow \text{AB} \rightarrow \text{AA} \rightarrow \text{SP}$ for S-based HS and $\text{AA}^\prime \rightarrow \text{AA} \rightarrow \text{AB}$ for Se-based HS.
This is marked by the IL (black dashed lines, $\mathrm{v}_1 - \mathrm{c}_1$ transition) and IN (red/blue dashed lines, $\mathrm{v}_2 - \mathrm{c}_2$ and $\mathrm{v}_1 - \mathrm{c}_4$, respectively) excitons. 
The AA and twisted stackings exhibit quite a different structure (as indicated by the higher total energy) and cannot be directly compared to the above.
The absorption spectra of MoSe$_2$/WSe$_2$ in the AA stacking [black continuous line, Fig.~\ref{fig:HSabspl4stackings}(b)] shows two absorption peaks for two intralayer excitons: one (red dashed line) localized on the MoSe$_2$ layer and the other (blue dashed line) on the WSe$_2$ layer. These transitions are independent but very close in energy. Both have significant and similar oscillator strengths (green triangles). The broadening chosen for visualization is the same for all plots, which, in this case, results in a merging of the two peaks into a single dome.
The vdW HS bright exciton peaks 
(Table~\ref{tab:infoexcHS}) are red shifted with respect to the constituent MLs (Table~\ref{tab:infoexcpeaksMLs}), except for the AA$^\prime$ (stable) and AB stacking of the S-based HS.
The BEs of the IN excitons are considerably smaller ($\sim$100 meV) in the vdW HS Table~\ref{tab:infoexcHS} with respect to the MLs Table~\ref{tab:infoexcpeaksMLs}. 
Thus, it is not straightforward to determine the layer localization of HS excitons simply comparing with the MLs. 

In the twisted HS the reduced interlayer distance and rotation compensate each other.
As a consequence, exciton positions are close to those of the MLs. 
Moreover, the IN-W exciton around$~\sim$1.8 eV is quenched, and it shows a small oscillator strength. We expect this to be a result of the compression along the stacking direction.

The magnitude of the $\Delta K_{\mathrm{Mo/W}}$ gaps varies among different stacking, the binding energy of the IL and IN excitons are close in magnitude (Table~\ref{tab:infoexcHS}), again except for the twisted HS which has a larger binding energy.
This is probably due to the fact that variations in the energy gaps and the contributions stemming from the many-body Kernel [as described in Eq.~(\ref{eq:BSE})] counterbalance one another.

\begin{figure}[th]
\includegraphics[width=\columnwidth]{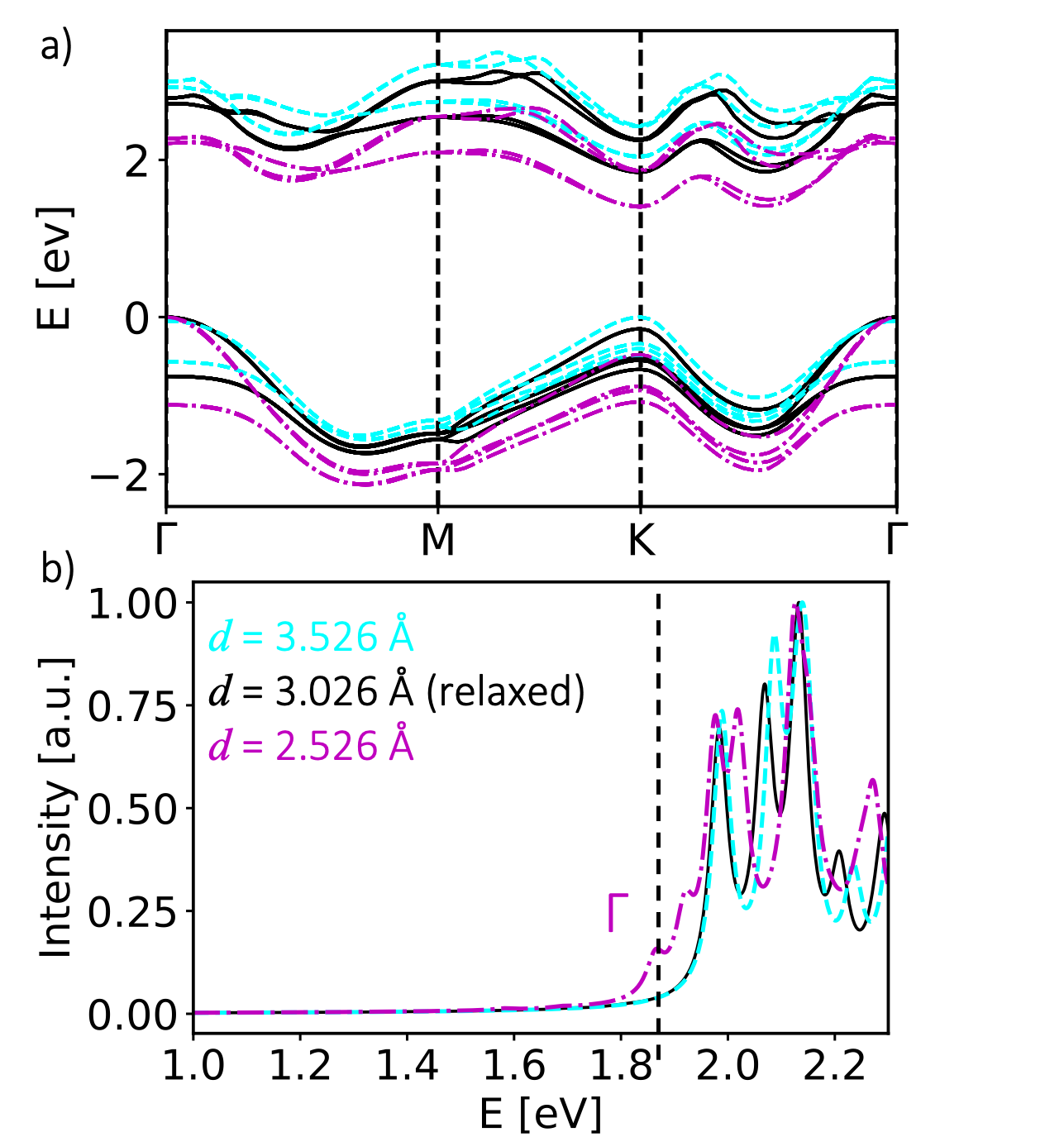}
    \caption{\label{fig:mos2ws2abvsdeltad} GW band structure (a) and BSE absorption spectrum (b) for MoS$_2$/WS$_2$ vdW HS in the AB stacking varying the interlayer distance.
    The black line corresponds to the  relaxed HS ($\Delta d = 0$), while cyan and magenta correspond to systems that differ in interlayer spacing by $\pm$0.5~\AA, respectively. Bands are aligned by their last valence state at $\Gamma$. The gap closes strongly with compression, and becomes direct under expansion.
    }
\end{figure}

\begin{figure}[ht]
    \includegraphics[width=0.9\columnwidth]{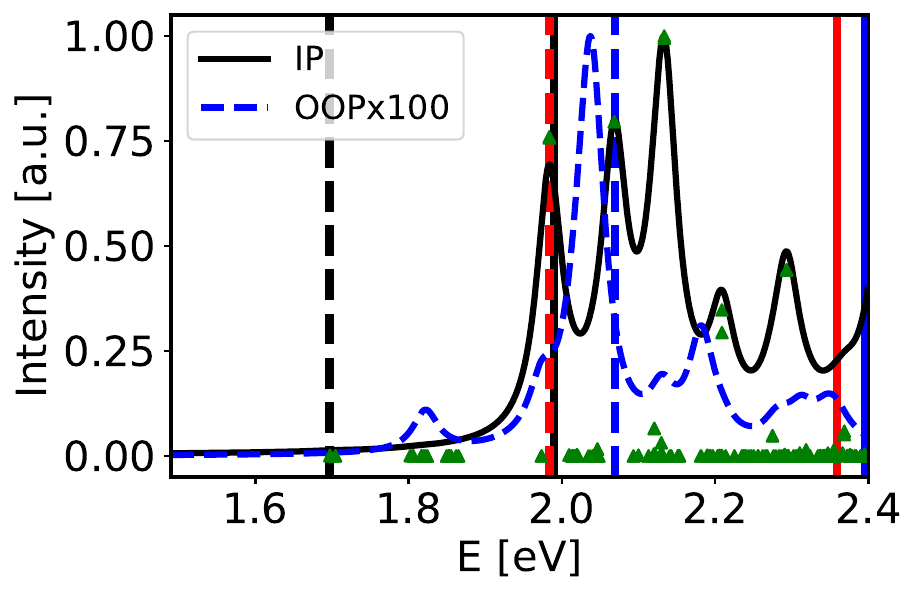}
    \caption{Absorption spectra of MoS$_2$/WS$_2$ (AB stacking) computed with in-plane (IP, black) or out-of-plane (OOP, blue) distribution of dipoles.
    The black continuous line indicates the 
    electronic gap, while the dashed lines indicate the exciton peak (Mo in red, W in blue). Green triangles represent the oscillator strength. Both spectra are rescaled to 1, but the OOP one is 100 times smaller than the IP.
    }
    \label{fig:mos2ws2ab_ipvsoop}
\end{figure}

\paragraph{Interlayer excitons:}
In all vdW HS (Fig.~\ref{fig:HSabspl4stackings}) we identify a dark and ``gray'' IL exciton states 
below the first IN-Mo and IN-W peaks.
Our results predict a reversed intralayer exciton
order, with respect to the monolayers, around $\sim$1.4 eV in the MoSe$_2$/WSe$_2$ AA$^\prime$ and AB stacking , with the IN-W exciton lower in energy than IN-Mo.
This unexpected reversed exciton order can be explained by the energy discrepancy in the electronic properties of the parent MLs (Sec.~\ref{sec:bandstructure}).
Note that the exciton order is very sensitive to the computational choices (SOC, pseudopotentials, van der Waals correction) employed for structural relaxation.

The electron distribution of the IL exciton in TMD HS does not reside only in the vacuum region between the two layers, contrary to  predictions based on continuum models. 
We understand this as a consequence of the nature of the exciton wavefunction, which formally contains contributions from all the $\mathbf{k}$-points in the Brillouin Zone. Such electron distribution cannot be  described in models that compute the optical properties only in proximity to the high-symmetry point $K$. 

Our results show that IL excitons of the twisted HS have a significant electron density weight in both layers (Table~\ref{tab:mose2wse2PIL}). Moreover, for the same system, we predict an exciton state just below the electronic gap, indicated in orange [Fig.~\ref{fig:HSabspl4stackings}(b)], featuring mixed intralayer MoSe$_2$ (v$_2$-c$_2$) and WSe$_2$ (v$_1$-c$_4$) transitions.
We label this state IN(Mo+W).
Analogously, not all the IN excitons have an electronic distribution highly localized within one layer (Tables~\ref{tab:mose2wse2PIL} and~\ref{tab:mos2ws2PIL}). An example of this can be seen in the AB stacking of MoSe$_2$/WSe$_2$, where the probability of IN localization, denoted as P(IN-W) [Eq.~(\ref{eq:INcoeff})], is only 20.21\%.

To assess the dependence of the exciton peaks on the interlayer distance $\Delta d$, we consider the specific case of MoS$_2$/WS$_2$ in the AB stacking (Fig.~\ref{fig:mos2ws2abvsdeltad}).
We varied $\Delta d$ by $\pm$ 0.5 \AA~ and inspected the effects on the band structure and absorption. Increasing $\Delta d$ the bands blueshift by about $\sim$200 meV, but absorption peaks blueshift by only $\sim$6 meV.  
This implies that c-axis tensile strain has a relatively weak effect on the position of the peaks.

Decreasing $\Delta d$ the bands redshift by $\sim$400 meV, the system is in an unfavourable configuration with high stresses and pressure. Absorption changes drastically, with new peaks arising between 1.8 and 1.9 eV [Fig.~\ref{fig:mos2ws2abvsdeltad}(b)].
In particular, the enhanced overlap between the $p_z$ orbitals due to vertical compression results in a new bright transition at $\Gamma$ at 1.86 eV.
The lack of inversion symmetry in vdW HS results in a $\sim$10 meV splitting of the c$_1$/c$_2$ bands. The effect of the splitting propagates to the absorption and emission spectra [Eq.~(\ref{eq:BSE})] where the first poles of the BSE equation with low oscillator strength [green triangles Figs~\ref{fig:HSabspl4stackings}(a) and~\ref{fig:HSabspl4stackings}(b)] are IL excitons, with energy splitting of the order of the SOC splitting~\cite{hanbicki2018double}.
This effect is more pronounced in Se-based HS due to Se larger atomic mass and SOC. 
In addition, MoS$_2$/WS$_2$ in the SP stacking displays two dark IL exciton state groups separated by 0.098 eV.
Se-based vdW HS, instead, display two distinct IL excitons in the AA$^\prime$ and AA stacking.
Furthermore, we observe that all systems have clusters of dark BSE equation poles below the first bright peak whose energy, shape, and spread
varies with the different stacking.

\paragraph{Dark and ``gray'' IL excitons:}
The darkness of excitons can be due to symmetry as in MLs but also to the interlayer nature, which implies very small overlap between the hole and electron wavefunctions. The latter excitons are ``gray'', but not dipole forbidden.
To discern the dark IL exciton, one can either inspect the dipole matrix elements (Table~\ref{tab:infoexcdipolesHSipvsoop}) or move the electric field polarization out of plane (schematically represented in Fig.~\ref{fig:schematic-ipvsoop}).
Considering again the AB alignment of MoS$_2$/WS$_2$ as a prototypical example, we see that varying the electric field direction, the forbidden IL exciton at $\sim$1.8 eV becomes bright, with the IN excitons being weaker than the in-plane (IP) configuration (Fig.~\ref{fig:mos2ws2ab_ipvsoop}).
In particular, we observe that, among the series of IL excitons below the fundamental gap, the first one remains dark in the out-of-plane (OOP) configuration while the second one becomes bright. This is due to different spin indices of the bands involved in the formation of the exciton states. The first IL exciton of the series comes from a transition between $\mathrm{v}_1-\mathrm{c}_1$ (spin-down/spin-up), while the second one from $\mathrm{v}_1-\mathrm{c}_2$ (spin-down/spin-down). 
Note that in Fig.~\ref{fig:mos2ws2ab_ipvsoop} we rescaled the maximum absorption to 1 for both IP and OOP configuration, however, in absolute value, the OOP one is around two orders of magnitude weaker due to the smaller out-of-plane dipole.

Detailed exciton energies and transitions for the various systems
are listed in Table~\ref{tab:infoexcHS}.
In Appendix \ref{app:hs} we report the absorption plots for IP and OOP electric field direction for all systems, along with information on the relevant energetic transitions.

\begin{table*}[t]
    \caption{\label{tab:infoexcHS} Interlayer exciton (IL), intralayer exciton (IN-M) and binding energies (BE) for the vdW HS.}
    \begin{tabular}{ccccc|ccc|cccc}
        \toprule
        \multirow{2}{*}{System} & \multirow{2}{*}{Stacking} & \multicolumn{3}{c}{IX$_0$ exciton} & \multicolumn{3}{c}{IX$_1$ exciton} & \multicolumn{3}{c}{IX$_2$ exciton} \\
        \cline{3-11}
         &  & E [eV] & BE [eV] & IL/IN & E [eV] & BE [eV] & IL/IN & E [eV] & BE [eV] & IL/IN  &  \\
        \midrule
        MoS$_2$/WS$_2$ & AA$^\prime$ & 1.765 & 0.292 & IL & 1.973 & 0.368 & IN-Mo & 2.079 & 0.323 & IN-W \\
         & AB & 1.697 & 0.293 & IL & 1.984 & 0.375 & IN-Mo & 2.069 & 0.327 & IN-W \\
         & SP & 1.383 & 0.290 & IL & 1.693 & 0.379 & IN-Mo & 1.786 & 0.332 & IN-W \\
         & AA & 1.515 & 0.283 & IL & 1.746 & 0.393 & IN-Mo & 1.827 & 0.345 & IN-W \\

        MoSe$_2$/WSe$_2$ & AA$^\prime$ & 1.205 & 0.222 & IL & 1.496 & 0.204 & IN-W & 1.565 & 0.175 & IN-Mo \\
        & AB & 1.094 & 0.235 & IL & 1.427 & 0.305 & IN-W & 1.478 & 0.275 & IN-Mo\\
         & AA & 1.198 & 0.252 & IL & 1.435 & 0.349 & IN-Mo & 1.587 & 0.164 & IN-W \\
         & 21.8$^\circ$ & 1.305 & 0.314 & IL & 1.613 & \multirow{2}{*}{\shortstack[l]{Mo: 0.308 \\ W: 0.257}}& IN-Mo and IN-W & 1.675 & 0.246 & IN-Mo \\
         &  &  &  &  & & &  &  & &  \\
         \bottomrule
    \end{tabular}
\end{table*}

\section{Photoluminescence and exciton radiative lifetimes}
Photoluminescence gives a quantitative description of the states involved in the creation of the 
IL exciton and their recombination time, which is crucial for any technological application, such as solar energy 
conversion devices~\cite{choi2021twist,lagarde2014carrier,korn2011low,shi2013exciton,peimyoo2013nonblinking}.

\begin{figure}[t]
    \includegraphics[width=\columnwidth]{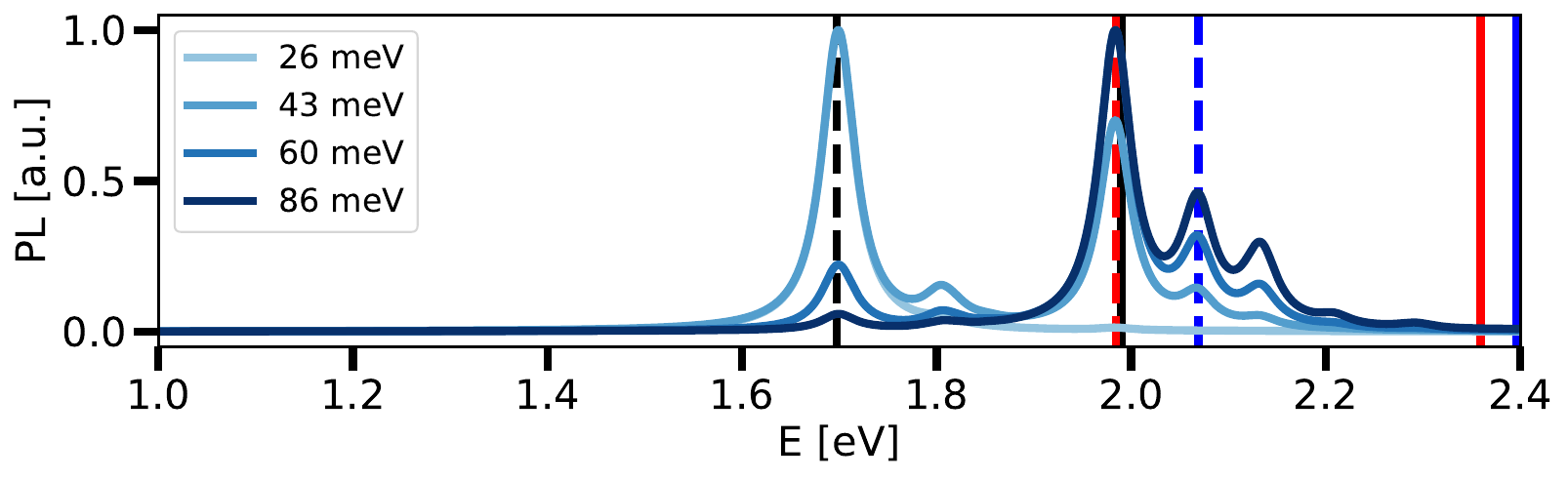}
    \caption{PL spectra of MoS2$_2$/WS$_2$ (AB stacking) as a function of the excitonic population $k_{\mathrm{B}}T_{\mathrm{exc}} = 26, 43, 60, 86$ meV. The vertical continuous and dashed lines are defined as in Fig.~\ref{fig:mos2ws2ab_ipvsoop}.
    The maximal PL intensity is rescaled to unity for each temperature.}
    \label{fig:mos2ws2abpl_T}
\end{figure}

Exciton population is tuned by changing the $T_{\mathrm{exc}}$ (in K or meV) in the Bose distribution.
For very small exciton population at $T_{\mathrm{exc}}=5$ K, the PL intensity peaks at the lowest dark exciton state.
In emission, we see that higher exciton states are populated and emit from the energy positions of the BSE poles (Fig.~\ref{fig:mos2ws2abpl_T}).

Fixing the exciton temperature $T_{\mathrm{exc}}$ at 400 K, we inspect the information carried by the PL spectra among different stacking (Fig.~\ref{fig:HSabspl4stackings}).
In analogy to the IP vs OOP electric field configuration described in Sec.~\ref{subsec:HSabspl}, the emission spectra show that not all dark IL excitons are the same in nature. Again, we can distinguish between dark (symmetry forbidden) and ``gray'' (small electron/hole overlap) IL excitons.
Interestingly, PL allows analyzing further the BSE poles, and it carries complementary information with respect to the absorption spectra (both IP and OOP).
Varying the direction of the electric field reveals the relative intensity of the exciton peaks in relation to the excitation direction. Meanwhile, PL serves as a distinct method to identify the most prominent interlayer (IL) emission peaks beneath the fundamental gap, among the series of BSE poles.
For example, in the AA$^\prime$ stacking (both S- and Se- based HS) we have two BSE poles around 1.77/1.19 eV, but only the second one is bright in emission.
The PL spectra for the Se-based 21.8$^\circ$ twisted HS is rather unique. IL excitons are not peaked at $K$ but between $\Gamma$ and $M$ (see Supplementary Information).
Note that, in addition to IL excitons, PL is peaked also for poles belonging to IN excitons.

The exciton radiative lifetimes can be modulated extensively across several
orders of magnitude by varying factors such as
the number of layers, stacking configurations, and temperature~\cite{chen2018theory,palummo2015exciton,re2021first,volmer2023twist}, as confirmed by our results illustrated in Fig.~\ref{fig:lifetimes}. Longer lifetimes are interesting, for instance, for photovoltaic applications~\cite{lagarde2014carrier}.

In MLs, the dark exciton states are long-lived ($\sim$1 Ms) attributable to 
the low exciton oscillator strength, suggesting a low probability
of re-emission upon occupation (Appendix Fig.~\ref{fig:MLslifetimes}).
Bright excitons possess lifetimes in the picosecond range, aligning with values reported in
similar studies~\cite{palummo2015exciton,chen2018theory}.
In vdW HS, we observe a significant change in the order of magnitude of the lifetimes depending on the stacking. 
The IL exciton lifetimes are in the pico/nanosecond range, with a couple of exceptional transitions in the microsecond range for the AA stacking.
The IN exciton lifetimes are all within the picosecond range, except for the mixed IN excitons of Se-based twisted vdW HS around 1.62 eV.
Moreover, the $21.8^\circ$ stacking shows a long-lived ground state IL exciton with lifetime $0.428 \mu \mathrm{s}$. 
The order of magnitude of the lifetime is inversely proportional to the exciton dipole strength [Eq.~(\ref{eq:tau_exc})].
IL excitons are characterized by an out-of-plane dipole moment. Its strength depends on the distance and especially on the overlap between the constituent electron and hole 
wave functions. A reduced overlap leads to a reduction of the exciton dipole strength and in turn an increase of its lifetime.
The order of magnitude of these excitons radiative lifetimes aligns quite well with available experiments which report, for the IL exciton lifetime, a value of $40 ps$ in MoS$_2$/WS$_2$ and $1.8$ ns in MoSe$_2$/WSe$_2$
~\cite{miller2017long,baranowski2017probing,kiemle2020control,rivera2015observation,jiang2021interlayer}.
To sum up, we show that there are IL excitons with ultralong lifetime of the order of $\mu s$ and that PL is a useful tool to identify stackings and investigate the nature of IL excitons in vdW HS.

\section{Conclusion}
We compute the electronic band structure, absorption spectra and photoluminescence for MoS$_2$/WS$_2$, MoSe$_2$/WSe$_2$ vdW heterostructures in different 
stacking configurations and compare to the constituent MLs. We discuss the shift and variations of the exciton peaks, order, and lifetimes 
with different stacking, and analyze the sensitivity of these results to the geometrical relaxation and
the inclusion of semicore states in the pseudopotentials.
We show that in vdW HS, the intralayer excitons have electron density spread in both layers. This feature is enhanced in the Se-based twisted HS where the mixed intralayer exciton has an electron density distributed over both MoSe$_2$ and WSe$_2$.
In addition, the twisted HS has a peculiar IL exciton with the electron density spread in both layers.
We discuss on the difference between spin-forbidden dark interlayer excitons and ``gray'' excitons, whose darkness is the result of a reduce spatial overlap between the electron and hole wavefunctions. We show that ``gray`` excitons become bright when the dipoles are excited in the stacking direction.
Furthermore, we show that the emission spectra obtained with our implementation of the PL spectra within the Ai-MBPT formalism in the \textsc{Yambo} code can bring additional insight into the nature of ``gray'' excitons.

Our findings highlight the potential to use absorption and emission as characterization tools for HS stacking.
The tunability of the excitonic properties of vdW TMDs heterostructures promises significant
applications in Quantum Information, especially in logical
qubit creation.

\begin{acknowledgments}
The authors acknowledge the fruitful discussion with Pedro M. M. C. de Melo, D. Vanmaekelbergh, M. Palummo, M. Re Fiorentin, A. Marini, and F. Paleari as well as extensive support from the \textsc{Yambo} developer team.
R.R., A.B.M. and Z.Z. acknowledge financial support from ``Materials for the Quantum Age -- QuMat'' project (Registration No. 024.005.006, Gravitation program of the Dutch Ministry of Education, Culture and Science OCW),
and the European Union ``Quondensate'' project (Horizon EIC Pathfinder Open, Grant Agreement No. 101130384).
RR and ZZ acknowledge financial support from Sector Plan Program 2019-2023.
MJV is supported by ARC project DREAMS (G.A. 21/25-11) funded by Federation Wallonie Bruxelles and ULiege, and the Excellence of Science (EOS) programme (grant 40007563-CONNECT) funded by the FWO and F.R.S.-FNRS.
The results of this research have been achieved using supercomputer facilities provided by NWO - Domain Science (Snellius),
the Tier-0 PRACE 
Research Infrastructure resource Discoverer based in Sofia, Bulgaria  (OptoSpin Project No. 2020225411) and Mare Nostrum by the Barcelona Supercomputing center 
(Spanish Supercomputing Network, RES Project No. FI-2020-1-0014). 

\end{acknowledgments}

\appendix
\section{The role of Se semicore states and van der Waals correction}\label{app:pseudos}
\renewcommand{\thefigure}{A\arabic{figure}}
\setcounter{figure}{0}

The treatment of the potential generated by the electronic states of Selenium (atomic number Z = 34) is critical for the assessment of the direct/indirect gap nature of TMDs MLs and, in general, of semiconductors and semimetals~\cite{rohlfing1998role}. In the pseudopotential approach, core electrons 
are considered to be frozen.
If only partial Se semicore states  
$3d^{10}4s^{2}4p^{4}$
are included in the pseudopotential, then we find an unphysical indirect GW band gap for the TMDs monolayers.
Including semicore $3s^{2}3p^{6}$ states and relaxing the structure with van der Waals corrections
provided the correct direct band structure. As an example, we provide a table of the relaxed lattice parameter $a$ for WSe$_2$ using 
different pseudopotentials, with/without van der Waals correction
(Table~\ref{table:WSe2-table}). 
In Fig.~\ref{fig:mose2wse2-spectrastdvsstr} we show these effects for the GW band structure of MoSe$_2$ and WSe$_2$. The shift in energy is also
reflected in the position of the exciton peaks in absorption. Coincidentally, the position of the first exciton for WSe$_2$ with partial semicore pseudopotentials is $\sim$1.63 eV  
and closer to the experimental value than the one reported in the main text, $~\sim$1.520 eV.
Moreover, employing partial semicore pseudopotentials, the MoSe$_2$ IX$_0$ exciton (at 1.52 eV) is lower in energy than the WSe$_2$ IX$_0$ exciton. 
Since full semicore pseudopotential lead to the correct band structure for the MLs, we employ full semicore pseudopotentials throughout this study for consistency.
We understand this difference as an increased screening due to the treatment of
the Coulomb potential as a truncated potential in a slab geometry~\cite{guandalini2023efficient}.

\begin{table}[h]
       \caption{Effects of Se semicore states and van der Waal correction in WSe$_2$ relaxed structural parameters (lattice parameter $a$ and thickness $\Delta t$), and GW electronic bandgap.} \label{table:WSe2-table}
       \begin{tabular}{c|c|c|c|c|c}
        \toprule
        a [\AA] & $\Delta t$ [\AA] & vdW & Semicore & $\Delta K$ gap [eV] & 
         GW gap  \\
        \hline
        3.323 & 3.360 & No & Partial  & 2.178 & Indirect \\
        3.323 & 3.360 & Yes & Partial  & 2.253 & Indirect\\
        3.338 & 3.251 & Yes & Full  & 1.934 & Direct \\
        \bottomrule
       \end{tabular}
\end{table}

\section{Structural properties of ML and HS}\label{app:tot-energies}
\renewcommand{\thefigure}{B\arabic{figure}}
\setcounter{figure}{0}

Relaxed structural parameters  (lattice parameter $a$, layer thickness $\Delta t$ and $\Delta t_{\mathrm{Mo/W}}$, and van der Waals gap $\Delta d$) of MLs and  HS are reported in Table \ref{tab:infogeometryHS}.

\begin{table}[h]
    \caption{\label{tab:infogeometryHS} Relaxed lattice parameter $a$, distance $\Delta d$ between layers, and layer thickness $\Delta t_{\rm{Mo/W}}$ for the vdW HS and TMDs MLs.}
    \begin{tabular}{c|c|c|c|c|c}
        \hline
        System & Stacking &$a$ [\AA] & $\Delta d$ [\AA]& $\Delta t_{\mathrm{Mo}}$ [\AA] & $\Delta t_{\mathrm{W}}$ [\AA] \\
        \hline
        MoS$_2$ & // & 3.194 & // & 3.129 & // \\
        WS$_2$ & // & 3.195 & // & // & 3.129  \\
        MoSe$_2$ & // & 3.317 & // & 3.301 & //  \\
        WSe$_2$ & // &3.338 & // & //& 3.251 \\
        MoS$_2$/WS$_2$ & AA$^\prime$ & 3.191 & 3.040 & 3.116 & 3.132 \\
        & AA & 3.191 & 3.633 & 3.117 & 3.131  \\
        & AB & 3.191 & 3.026 & 3.116 & 3.130 \\
        & SP & 3.191 & 3.523 & 3.116 & 3.131 \\
        MoSe$_2$/WSe$_2$ & AA$^\prime$ & 3.327 & 3.137 & 3.329 & 3.342 \\
        & AA & 3.327  & 3.721 & 3.328 & 3.342  \\
        & AB & 3.327 & 3.138 & 3.328 & 3.342 \\
        & SP & 3.327 & 3.253 & 3.328 & 3.343  \\
        & 21.8$^\circ$ & 8.780 & 2.954 & 3.280 & 3.333 \\      
        \hline  
    \end{tabular}
    \end{table}

\section{GW band structures for TMD HS}\label{app:GW}
\renewcommand{\thefigure}{C\arabic{figure}}
\setcounter{figure}{0}

\begin{figure}[t]
    \centering
    \includegraphics[width=0.9\columnwidth]{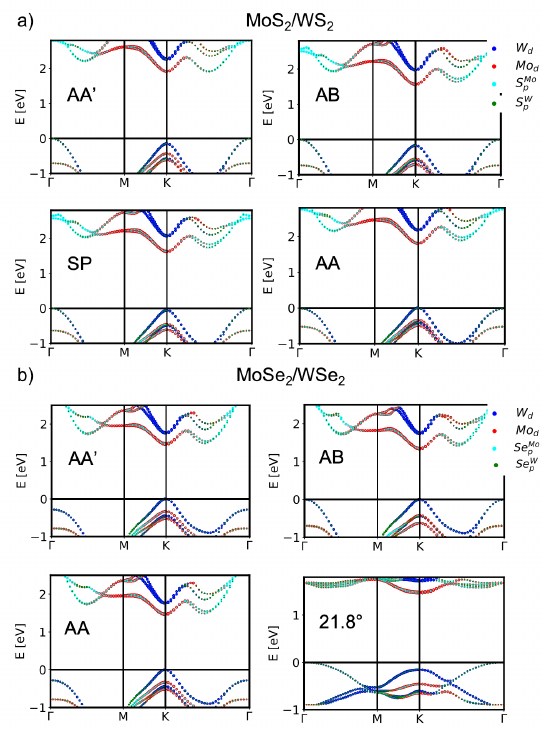}
    \caption{Orbital projected GW band structure of MoS$_2$/WS$_2$ (a) and MoSe$_2$/WSe$_2$ (b) for different stackings. The size of the dots is weighted by the relative orbital contribution.} 
       \label{fig:moX2wX2-projbands}
\end{figure}

For each system, we report the orbital-projected band structure in Fig.~\ref{fig:moX2wX2-projbands}. 
Our findings indicate that all systems exhibit a type II band alignment, with the primary composition of the first four conduction and valence bands being the d-orbitals of the metals and the p-orbital of the chalcogen atom.
Although the orbital projections vary only slightly among the systems, the $\Delta KQ$ and the $\Delta K\Gamma$
splittings are strongly influenced by the different stacking configurations, as reported in the main text and in Table~\ref{tab:infoHSsSOC}.

\begin{table*}[h]
\caption{\label{tab:infoHSsSOC} GW Energy differences $\Delta K$, $\Delta K_{\mathrm{Mo}}$ and $\Delta K_{\mathrm{W}}$, energy valley offsets ($\Delta K\Gamma = E_{K}-E_{\Gamma}$, $\Delta KQ = E_{Q}-E_{K} $) and spin-orbit coupling splitting at $K$ for different stacking of MoS$_2$/WS$_2$ and MoSe$_2$/WSe$_2$ (all in eV).
}
\begin{tabular}{c|c|c|c|c|c|c|c|c|c|c|c|c}
\hline
System & stacking & $\Delta K$ & $\Delta K_{\mathrm{Mo}}$ & $\Delta K_{\mathrm{W}}$ & $\Delta K\Gamma$ & $\Delta KQ$ & $\Delta K_{\mathrm{v}_1\mathrm{v}_2}$ & $\Delta K_{\mathrm{v}_2\mathrm{v}_3}$ & $\Delta K_{\mathrm{v}_3\mathrm{v}_4}$ & $\Delta K_{\mathrm{c}_2\mathrm{c}_1}$ & $\Delta K_{\mathrm{c}_3\mathrm{c}_2}$ & $\Delta K_{\mathrm{c}_4\mathrm{c}_3}$ \\
\hline
MoS$_2$/WS$_2$ & AA$^\prime$ & 2.058 & 2.341 & 2.403 & --0.150 & 0.000 & 0.283 & 0.142 & 0.033 & 0.008 & 0.336 & 0.015\\
& AB & 1.990 & 2.359 & 2.397 & --0.149 & 0.009 & 0.369 & 0.142 & 0.117 & 0.008 & 0.336 & 0.011 \\
& SP & 1.673 & 2.072 & 2.118 & --0.060 & 0.122 & 0.398 & 0.033 & 0.124 & 0.009 & 0.435 & 0.011 \\
& AA & 1.798 & 2.139 & 2.172 & 0.021 & 0.094 & 0.341 & 0.066 & 0.086 & 0.010 & 0.365 & 0.008\\
MoSe$_2$/WSe$_2$ & AA$^\prime$ & 1.427 & 1.748 & 1.706 & 0.028 & 0.030 & 0.331 & 0.018 & 0.079 & 0.027 & 0.383 & 0.011 \\
&AB & 1.329 & 1.753 & 1.731 & 0.006 & 0.090 & 0.424 & 0.018 & 0.183 & 0.019 & 0.383 & 0.025 \\
&AA & 1.452 & 1.784 & 1.751 & 0.281 & 0.030 & 0.331 & 0.113 & 0.079 & 0.027 & 0.272 & 0.011\\
&21.8$^\circ$ & 1.619 & 1.921 & 1.870 & --0.156 & 0.159 & 0.302 & 0.150 & 0.045 & 0.025 & 0.226 & 0.015 \\
\hline
\end{tabular}
\end{table*}

\begin{figure}[t]
\includegraphics[width=\columnwidth]{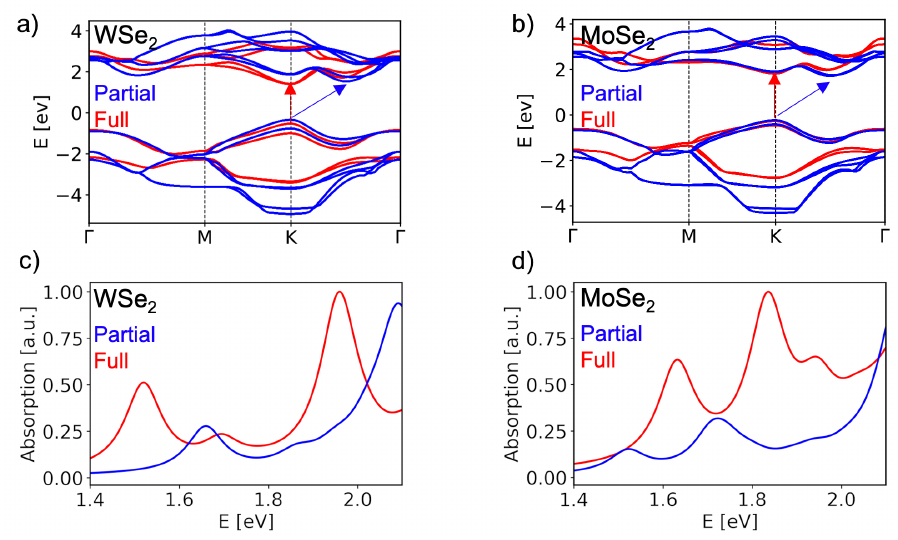}
\caption{GW (a,b) band structures and absorption (c,d) spectra for WSe$_2$ and MoSe$_2$ MLs with full semicore (red) or partial semicore (blue) pseudopotentials. The inclusion of semicore states and relaxing the structures with SOC correctly give a direct gap band structure.}
\label{fig:mose2wse2-spectrastdvsstr} 
\end{figure}

\section{Absorption and Emission Spectra of TMD Monolayers}\label{app:mlsabsemission}
\renewcommand{\thefigure}{D\arabic{figure}}
\setcounter{figure}{0}
The absorption and emission spectra [Eq.~(\ref{eq:BSE}) and Eq.~(\ref{eq:polarizability})] of the MLs (Fig.~\ref{fig:MLsabsploscstrength}), confirm the high exciton
binding energy attributed to diminished dielectric screening in the 2D limit.
We observe a prominent emission peak at the first 
exciton eigenstate of the BSE Hamiltonian [Eq.~(\ref{eq:BSE})]. 
This peak coincides with the position of the first finite oscillator strength (green triangles) and is dark in absorption due to optical selection rules (spin forbidden).
MoSe$_2$ has a dark exciton degenerate with the first 
bright peak, confirming previous reports~\cite{wang2018colloquium,yu2015valley}.
The computed excitonic peaks (labeled as IX$_N$, where $N=1,2...$), align closely with experimental data 
for the S-based MLs~\cite{hong2014ultrafast} (Table~\ref{tab:infoexcpeaksMLs}).
The results for Se-based MLs are further from experimental findings~\cite{wilson2017determination} (e.g. the energy of the first exciton peak in WSe$_2$ is about 100--200 meV 
lower
) due to the underestimation of the electronic energy gap (Sec.~\ref{sec:bandstructure}).

\begin{table}[h]
    \caption{\label{tab:infosplittingsMLs}Energy valley splittings and GW gap for the MLs.}
    \begin{tabular}{c|c|c|c}
    \hline
     ML & $\Delta K$ gap [eV] & $\Delta K\Gamma$ [eV] & $\Delta KQ$ [eV]  \\
    \hline
     MoS$_2$ & 2.397 & 0.122 & 0.184  \\
     WS$_2$ & 2.462 & 0.252 & 0.028 \\
     MoSe$_2$ & 2.089 & 0.345 & 0.144 \\
     WSe$_2$ & 1.934 & 0.312 & 0.309  \\
     \hline
    \end{tabular}
\end{table}

\begin{figure}[ht]
    \includegraphics[width=0.5\textwidth]{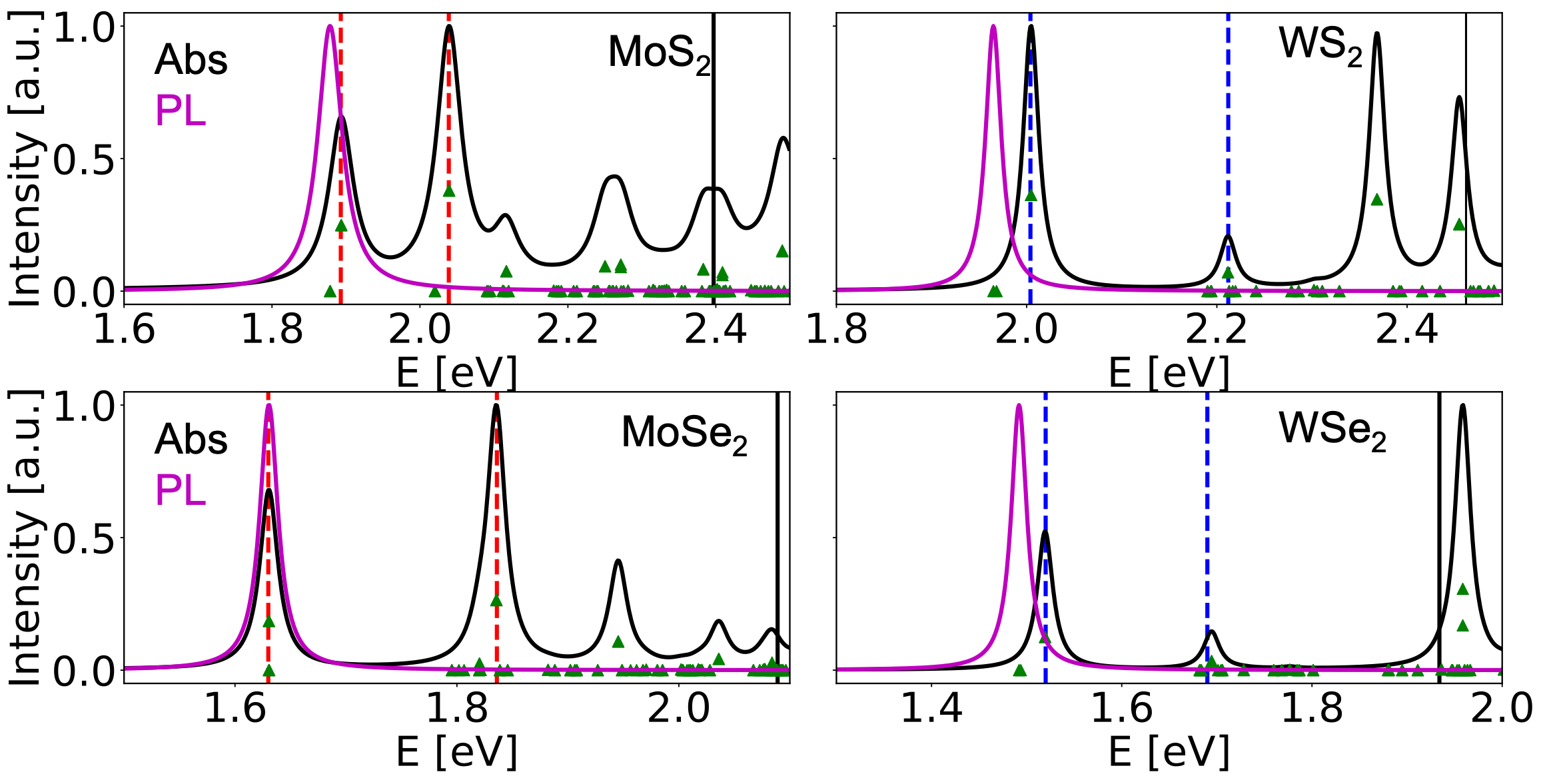}
    \caption{Absorption (black) and PL(magenta, $T_{\mathrm{exc}}=5 K$) plots for different MLs. The black continuous vertical lines indicate the 
    GW electronic gap while the dashed vertical lines indicate the excitonic peaks (Mo in red, W in blue). Green triangles represent the oscillator strength.}
    \label{fig:MLsabsploscstrength}
\end{figure}

\begin{figure}[h]
    \includegraphics[width=0.48\textwidth]{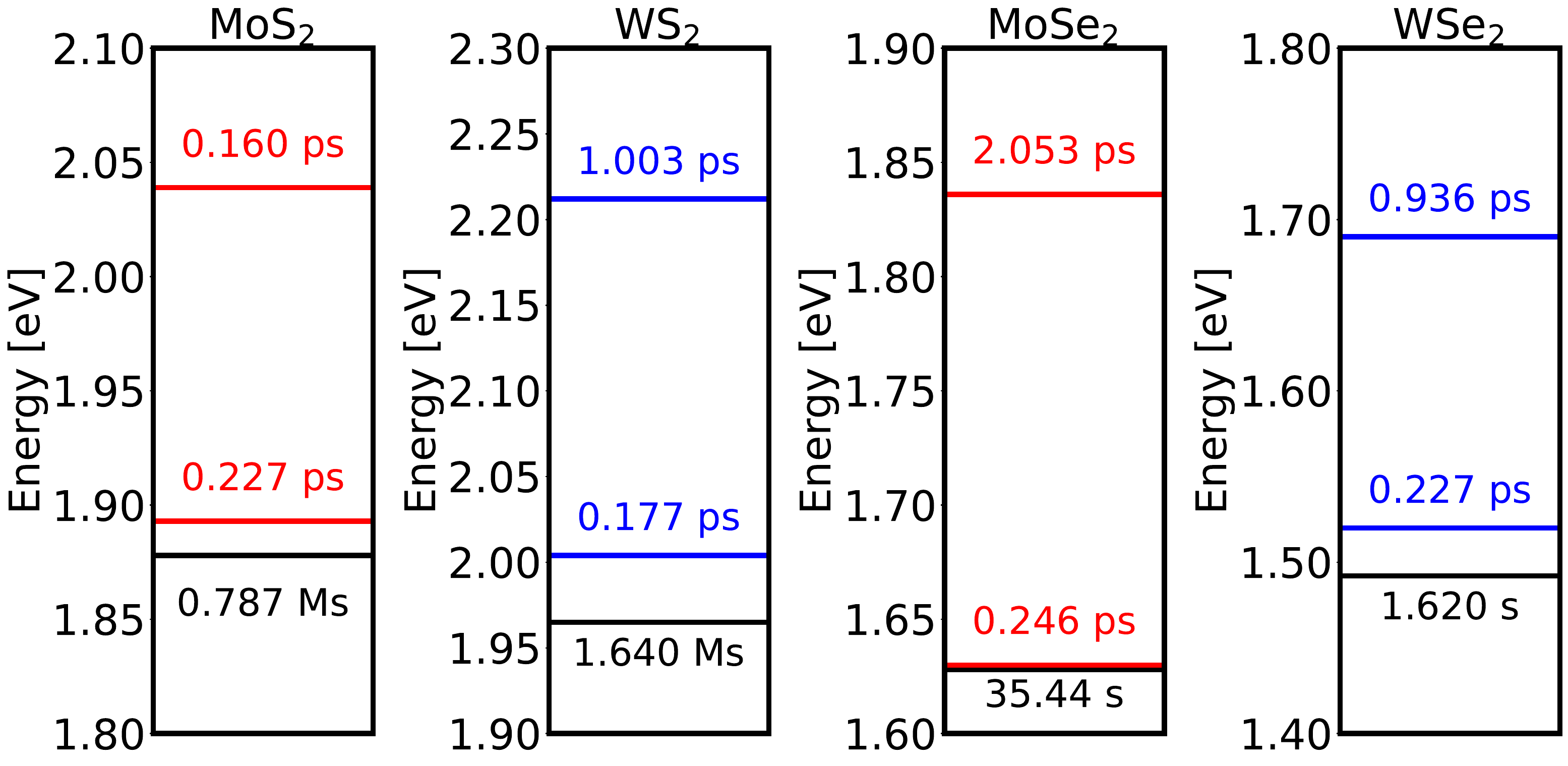} Here is how to import EPS art
    \caption{\label{fig:MLslifetimes}Lifetimes of the monolayers for IL dark states (black line) and IN-Mo/IN-W bright states (red and blue lines).
    }
\end{figure}

\section{Optical absorption of TMD heterostructures}\label{app:hs}
\renewcommand{\thefigure}{E\arabic{figure}}
\setcounter{figure}{0}
We present a comprehensive table (Table~\ref{tab:infoexcHSipvsoop}) detailing the most significant transitions observed in the BSE spectra, which manifest in both absorption (IP and OOP) and photoluminescence (PL). Fig.~\ref{fig:schematic-ipvsoop} shows a schematic of the IP and OOP setup. Fig.~\ref{fig:HSabsipvsoop} displays the IP and OOP absorption spectra for the systems under investigation, and the normalized exciton dipoles are reported in Table~\ref{tab:infoexcdipolesHSipvsoop}.

Analysis indicates that the strongest peaks in the OOP configuration do not appear in the IP configuration, and the most prominent IP peaks are diminished in the OOP configuration. To elucidate these findings, we have employed dashed and continuous lines in Figs.~\ref{fig:HSabsipvsoop} and \ref{fig:HSabspl4stackings} to denote relevant excitonic transitions, thereby underscoring their significance within the scope of this discussion.
To quantify the differences between ``dark'' excitons (dark in both IP and OOP configurations) and ``gray'' excitons (dark in the IP configuration but bright in the OOP configuration, or vice versa), we present the excitonic dipole strengths of selected transitions in Table~\ref{tab:infoexcdipolesHSipvsoop}. 

In particular, Fig.~\ref{fig:HSabspl4stackings}(b) highlights various excitonic features in the twisted heterostructure: interlayer (IL) excitons are marked with dashed vertical black lines, the mixed intralayer IN(Mo+W) state is indicated in orange, and the first three intralayer IN-Mo excitonic states are also delineated. Notably, no bright intralayer W (IN-W) excitons are observed. Additionally, Fig.~\ref{fig:HSabsipvsoop}(b) introduces new transitions following the same color coding, with an observable IN-W exciton in the twisted configuration at 1.578 eV.
\begin{figure}[ht]
    \includegraphics[width=0.5\textwidth]{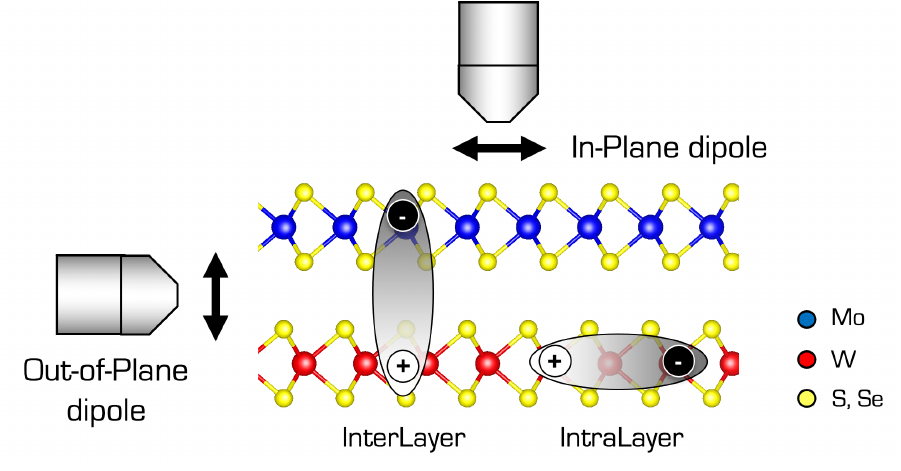}
    \caption{Schematic of a prototypical experimental setup setup showing a detector (gray) capturing either in-plane (IP) or out-of-plane (OOP) light excitations. The ball and stick model depicts a Se-based vdW HS.}
    \label{fig:schematic-ipvsoop}
\end{figure}

\begin{table*}[h]
\caption{\label{tab:mos2ws2PIL}GW gap, exciton, and binding energies (in eV) and IN/IL coefficients for different stackings of MoS$_2$/WS$_2$.}
\begin{tabular}{c|c|c|c|c|c|c|c}
\toprule
 Stacking&$\Delta K$ [eV]&IX0 & P(IL) &  IX1 & P(IN-Mo) & IX2 & P(IN-W)\\
\cline{1-8}
 AA$^\prime$ &1.960 &1.670 (0.290) & 97.81\% & 1.889 (0.366) & 94.06\% & 2.057 (0.255) & 54.74\%\\
 AA & 1.798 &1.515 (0.283)& 99.2\% & 1.746 (0.393)& 99.7\% & 1.827 (0.345) & 65.24\%\\
 AB & 1.727 & 1.194 (0.533) & 99.2\% & 1.496 (0.618) & 98.3\%& 1.639 (0.475) & 99.0\%\\
 SP &1.673 &1.383 (0.290) &99.14\% &1.693 (0.379) &80.15\% & 1.786 (0.332) &97.26\%\\
\bottomrule
\end{tabular}
\end{table*}

\begin{table*}[h]
\caption{\label{tab:mose2wse2PIL}GW gap, exciton and binding energies (in eV) and IL/IN coefficient for different stackings of MoSe$_2$/WSe$_2$.}
\begin{tabular}{c|c|c|c|c|c|c|c}
 \toprule
 Stacking&$\Delta K$ [eV]&IX0 & P(IL) &  IX1 & P(IN-) & IX2 & P(IN-)\\
 \cline{1-8}
 AA$^\prime$ &1.427 &1.205(0.222) & 97.70\%&1.496(0.204) & 57.91\%(W)& 1.565(0.175) & 84.38\% (Mo)\\
 AA & 1.452 &1.198(0.252)& 96.98\%&1.435(0.349)& 99.6\%(Mo)&1.587(0.164) & 59.06\% (W)\\
 AB & 1.329 &1.094(0.235)& 98.71\%&1.427(0.3049)& 20.21\%(W)&1.478(0.275) &86.16\% (Mo)\\
 21.8$^\circ$ & 1.619 & 1.305(0.314) & 52.63\%& 1.613(0.308/0.257) & 40.31\%(W)& 1.675(0.246) & 25.63\%(Mo)\\
\bottomrule
\end{tabular}
\end{table*}

\begin{table}[h]
    \caption{\label{tab:infoexcpeaksMLs} Exciton (IX$_N$), PL peaks and binding energy (BE) for the different MLs.}
    \begin{tabular}{c|c|c|c|c|c}
    \hline
     ML & IX$_1$ [eV] & BE [eV] & IX$_2$ [eV] & BE [eV] & PL [eV] \\
     \hline
     MoS$_2$ & 1.893 & 0.504 & 2.039 & 0.358 & 1.878 \\
     WS$_2$ & 2.004 & 0.458 & 2.212 & 0.250 & 1.965\\
     MoSe$_2$ & 1.630 & 0.459 & 1.836 & 0.253 & 1.630 \\
     WSe$_2$ & 1.520 & 0.414 & 1.694 & 0.240 & 1.492 \\
     \hline
    \end{tabular}
\end{table}

\begin{table*}[ht]
    \centering
    \caption{\label{tab:infoexcHSipvsoop} Transition table for Interlayer exciton (IL) and intralayer exciton (IN-X) for all vdW HSs. We report the pronounced peaks for 
    IP and OOP electric field direction depicted in Fig.~\ref{fig:HSabsipvsoop}. A slash symbol in the binding energy indicates that the transition is above the 
    electronic gap.}
    \begin{tabular}{ccc|ccc}
        \toprule
        \multicolumn{3}{c}{MoS$_2$/WS$_2$ AA$^\prime$} & \multicolumn{3}{c}{MoSe$_2$/WSe$_2$ AA$^\prime$} \\
        \cline{1-6}
        \midrule
        E [eV] & BE [eV] & Transition ($K$ point) & E [eV] & BE [eV] & Transition ($K$ point) \\
        1.765  & 0.292   & IL ($K$) v$_1$-c$_1$ & 1.184 & 0.243 & IL($K$) v$_1$-c$_1$ \\
        1.771  & 0.298   & IL ($K$) v$_1$-c$_1$ & 1.205 & 0.222 & IL($K$) v$_1$-c$_2$ \\
        1.867 & 0.190 & IL ($K$) v$_1$-c$_1$ & 1.205 & - & - \\
        1.964  & 0.377   & IN-Mo ($K$) v$_2$-c$_1$ & 1.275 & 0.152 & IL ($\Gamma-K$, close to $K$) v$_1$-c$_1$ \\
        -  & -   & - & 1.290 & 0.169 & IL ($\Gamma-K$, close to $K$) v$_1$-c$_2$ \\        
        1.973  & 0.368   & IN-Mo ($K$) v$_2$-c$_1$ &  1.462 & /  & IL ($\Gamma-K$ close to $K$) v$_1$-c$_2$ \\
        2.043  & 0.359   & IN-W ($K$) v$_1$-c$_4$  & 1.496 & 0.204 & IN-W ($K$) v$_1$-c$_4$ \\
        2.079  & 0.323   & IN-W ($K$)     v$_1$-c$_4$  & 1.565 & 0.175 & IN-Mo ($K$) v$_2$-c$_1$ \\        
        \cline{1-6}
        \multicolumn{3}{c}{MoS$_2$/WS$_2$ AB} & \multicolumn{3}{c}{MoSe$_2$/WSe$_2$ AB} \\
        \cline{1-6}
        E [eV] & BE [eV] & Transition ($K$ point) & E [eV] & BE [eV] & Transition ($K$ point)\\
        1.697  & 0.293   & IL ($K$) v$_1$-c$_1$ &  1.094 & 0.235 & IL($K$) v$_1$-c$_1$ \\
          -    &   -     &   -                  &  1.187 & 0.142 & IL ($\Gamma-K$, close to $K$) v$_1$-c$_1$\\
          -    &   -     &   -                  &  1.332 & - & IL ($\Gamma-K$, close to $K$) v$_1$-c$_1$\\
        1.822  & 0.537  & IL ($K$) v$_1$-c$_2$ & 1.367 & - & IL ($\Gamma-K$, close to $K$) v$_1$-c$_1$  \\
        1.984  & 0.375   & IN-Mo ($K$) v$_2$-c$_1$ & 1.420 & - & IL ($M-K$, close to $K$) v$_1$-c$_1$\\
        2.069  & 0.327   & IN-W ($K$) v$_1$-c$_4$ & 1.427 & 0.305 & IN-W ($K$) v$_1$-c$_4$\\          
        2.120  &    /    & IL ($\Gamma-K$, close to $K$) v$_1$-c$_1$ & 1.478 & 0.275 & IN-Mo ($K$) v$_2$-c$_2$ \\
        \cline{1-6}
        \multicolumn{3}{c}{MoS$_2$/WS$_2$ SP} & \multicolumn{3}{c}{MoSe$_2$/WSe$_2$ AA} \\
        \cline{1-6}
        E [eV] & BE [eV] & Transition ($K$ point) & E [eV] & BE [eV] & Transition ($K$ point) \\
        1.383  & 0.290   & IL ($K$) v$_1$-c$_1$ & 1.198 & 0.252 & IL ($K$) v$_1$-c$_1$\\
        -  & -   & - & 1.223 & 0.278 & IL ($K$) v$_1$-c$_2$\\
        1.481  & 0.192   & IL ($K$) v$_1$-c$_1$ & 1.285 & 0.168  & IL ($K$) v$_1$ - c$_1$ \\
        1.684  &   /  & IL ($K$) v$_1$-c$_2$ & 1.435 & 0.349 & IN-Mo ($K$) v$_2$-c$_1$ \\
        1.693  & 0.379   & IN-Mo ($K$) v$_2$-c$_1$ & 1.453 & 0.298 & IN-W ($K$) v$_1$-c$_4$ \\
          -    &  -     &   -                     & 1.554 &  /      & IL ($\Gamma-K$, close to $K$) v$_1$-c$_1$\\
        1.753  & 0.367   & IN-W ($K$) v$_1$-c$_3$ & 1.587 & 0.164 & IN-W ($K$) v$_1$-c$_4$ \\  
        \cline{1-6}        
        \multicolumn{3}{c}{MoS$_2$/WS$_2$ AA} & \multicolumn{3}{c}{MoSe$_2$/WSe$_2$ 21.8$^\circ$} \\
        \cline{1-6}
        E [eV] & BE [eV] & Transition ($K$ point) & E [eV] & BE [eV] & Transition ($K$ point) \\
        1.515  & 0.283   & IL ($K$) v$_1$-c$_1$ & 1.305 & 0.314 & IL ($\Gamma$ and $\Gamma-M$, close to $\Gamma$) v$_1$-c$_1$\\  
        -  & -   & - & 1.332 & 0.287 & IL (between $\Gamma$ and $\Gamma-M$) v$_1$-c$_1$\\             
        1.629  & 0.169   & IL ($K$) v$_1$-c$_2$ &   1.403 & 0.216 & IL ($\Gamma$) v$_1$-c$_1$\\
        1.735  & 0.063   & IL ($\Gamma-K$, close to $K$ v$_1$-c$_1$) &  1.578 & 0.292 & IN-W ($K$) v$_1$-c$_3$ \\
        1.746  & 0.393  & IN-Mo ($K$) v$_2$-c$_1$ & 1.613 & \multirow{2}{*}{\shortstack[l]{Mo: 0.308 \\ W: 0257}} & \multirow{2}{*}{\shortstack[l]{IN-Mo ($K$) v$_2$-c$_1$ \\ IN-W ($K$): v$_1$-c$_4$}}  \\\\
         -    &  -     &   -                     & 1.675 & 0.246 & IN-Mo ($K$) v$_2$-c$_1$  \\          
        1.827  & 0.345   & IN-W ($K$) v$_1$-c$_4$ & 1.720 & 0.150 & IN-W ($\Gamma-K$, close to $K$) v$_1$-c$_4$ \\   
        \bottomrule
    \end{tabular}
\end{table*} 

\begin{table}[ht]
    \centering
    \caption{\label{tab:infoexcdipolesHSipvsoop} In plane (IP) and out of plane (OOP) excitonic dipoles $d$ (see also Fig.~\ref{fig:HSabsipvsoop}) for some of the interlayer excitons (IL) and intralayer excitons (IN-X) for all vdW HSs. We differentiate between dark excitons, with $d=0$ (we set to zero anything smaller than $5\times10^{-5}$), ``gray'' with $5\times10^{-5}<d<0.1$, and bright excitons with $d\geq 0.1$. The values are w.r.t. the most intense excitonic dipole which is set to 1.
    We highlight ``gray`` excitons in bold.
    Excitonic dipoles are normalized to 1 relative to the largest dipole in the system. A slash in the binding energy column indicates that the transition exceeds the electronic gap.
    }
    \begin{tabular}{ccc|ccc}
        \toprule
        \multicolumn{3}{c}{MoS$_2$/WS$_2$ AA$^\prime$} & \multicolumn{3}{c}{MoSe$_2$/WSe$_2$ AA$^\prime$} \\
        \cline{1-6}
        \midrule
        E [eV] & IP & OOP & E [eV]  & IP & OOP \\
        \cline{1-6}
        \textbf{1.765}  & 2.55$\times10^{-4}$ & 0.  & \textbf{1.184} & 1.83$\times10^{-4}$ & 0. \\
        1.964  & 0. & .162 & \textbf{1.275} & 0. & 1.07$\times10^{-2}$ \\
        1.973  & .802 & 0.  & - & - & - \\
        2.043  & 0. & 1.0 & 1.496 & .255 & 0. \\
           -   &  -  & -   & 1.558 & 0. & 1.0 \\
        2.079  & .882 & 0. & 1.565 & .562 & 0. \\  
        \cline{1-6}
        \multicolumn{3}{c}{MoS$_2$/WS$_2$ AB} & \multicolumn{3}{c}{MoSe$_2$/WSe$_2$ AB} \\
        \cline{1-6}
        \textbf{1.697}  & 1.68$\times10^{-3}$ &0.  & \textbf{1.094} & 2.12$\times10^{-3}$ & 0. \\
          -    & - & - & 1.187 & 0. & .160\\
          -    & - & - & 1.332 & 0. & .248 \\
        1.822  &  0. & .135 & 1.367 & 0. & .331 \\
        1.984  &  .76 & 0. & 1.420 & 5.82$\times10^{-5}$ & .514 \\
        2.069  & .80 & 0. & 1.427 & .416 & 0. \\          
        \textbf{2.122}  & 0. & 1.81$\times10^{-3}$  & 1.478 & 0. & 0. \\
        \cline{1-6}
        \multicolumn{3}{c}{MoS$_2$/WS$_2$ SP} & \multicolumn{3}{c}{MoSe$_2$/WSe$_2$ AA} \\
        \cline{1-6}
        \textbf{1.383}  & 3.68$\times10^{-4}$ & 4.06$\times10^{-3}$  & \textbf{1.198} & 5.78$\times10^{-3}$ & 0. \\
        \textbf{1.481}  & 2.23$\times10^{-3}$ & 4.49$\times10^{-2}$  & \textbf{1.285} & 0. & 2.42$\times10^{-2}$ \\
        1.684  & 1.67$\times10^{-4}$ & .13  & \textbf{1.435} & 8.5$\times10^{-2}$\\
        1.693  & .64 & 6.39$\times10^{-4}$  & 1.453 & .96 & 0. \\
        1.753  & 0. & 1.0  & 1.587 & .97 & 0.\\  
        1.786  &  .64 & 0. &  -   & - & - \\
        \cline{1-6}        
        \multicolumn{3}{c}{MoS$_2$/WS$_2$ AA} & \multicolumn{3}{c}{MoSe$_2$/WSe$_2$ 21.8$^\circ$} \\
        \cline{1-6}
        \textbf{1.515}  & 3.42$\times10^{-3}$& 0.  & \textbf{1.332} & 0. & 8.56$\times10^{-2}$ \\             
        \textbf{1.629}  & 0. & 1.00$\times10^{-2}$  & 1.578 & 0. & 1.00 \\
        \textbf{1.735}  & 0. & 1.26$\times10^{-3}$  &1.613 & 0.14 & 0. \\
        1.746  & 1.00 & 0.  & 1.675 & 0.74 & 0. \\        
        1.827  & 0.92 & 0.  & \textbf{1.720} & 0. & 5.39$\times10^{-2}$ \\   
        \bottomrule
    \end{tabular}
\end{table} 

\begin{figure*}[t]
    \centering
    \includegraphics{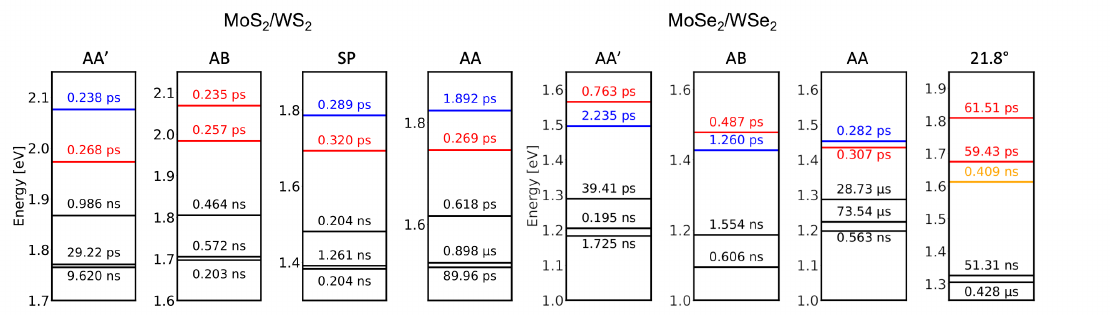}
    \caption{Energies and lifetimes of the vdW HS for dark IL states (black line) and bright IN-Mo/IN-W states (red and blue lines).
    The orange line denotes the mixed IN+Mo/W state for the $21.8^\circ$ twist stacking.}
    \label{fig:lifetimes}
\end{figure*}

\begin{figure*}[h]
   \centering
   \includegraphics[width=1.0\textwidth]{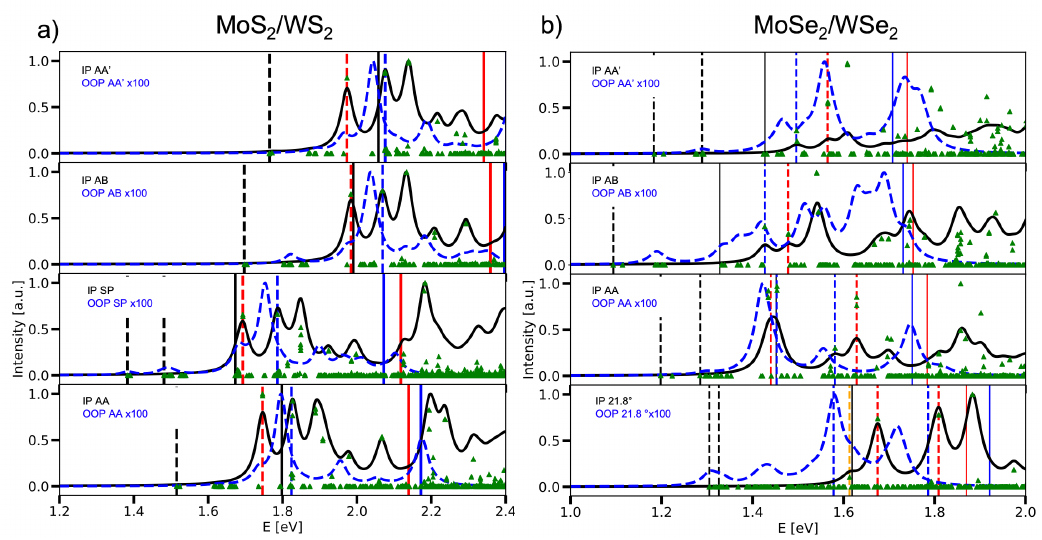}
   \caption{Absorption spectra for (a) MoS$_2$/WS$_2$ and (b) MoSe$_2$/WSe$_2$ HS, computed IP or OOP distribution of dipoles. Oscillator strengths are represented by green triangles. Dashed lines in black, red, and blue highlight the IP transition character at the $K$ point, corresponding to IL, IN-Mo, IN-W transitions, respectively. Vertical solid lines in black, red, and blue delineate the electronic bandgap along with gaps between Mo and W projected bands at $K$.}
   \label{fig:HSabsipvsoop}
\end{figure*}

\newpage
\bibliography{bibliography}

\begin{thebibliography}{61}%
\makeatletter
\providecommand \@ifxundefined [1]{%
 \@ifx{#1\undefined}
}%
\providecommand \@ifnum [1]{%
 \ifnum #1\expandafter \@firstoftwo
 \else \expandafter \@secondoftwo
 \fi
}%
\providecommand \@ifx [1]{%
 \ifx #1\expandafter \@firstoftwo
 \else \expandafter \@secondoftwo
 \fi
}%
\providecommand \natexlab [1]{#1}%
\providecommand \enquote  [1]{``#1''}%
\providecommand \bibnamefont  [1]{#1}%
\providecommand \bibfnamefont [1]{#1}%
\providecommand \citenamefont [1]{#1}%
\providecommand \href@noop [0]{\@secondoftwo}%
\providecommand \href [0]{\begingroup \@sanitize@url \@href}%
\providecommand \@href[1]{\@@startlink{#1}\@@href}%
\providecommand \@@href[1]{\endgroup#1\@@endlink}%
\providecommand \@sanitize@url [0]{\catcode `\\12\catcode `\$12\catcode `\&12\catcode `\#12\catcode `\^12\catcode `\_12\catcode `\%12\relax}%
\providecommand \@@startlink[1]{}%
\providecommand \@@endlink[0]{}%
\providecommand \url  [0]{\begingroup\@sanitize@url \@url }%
\providecommand \@url [1]{\endgroup\@href {#1}{\urlprefix }}%
\providecommand \urlprefix  [0]{URL }%
\providecommand \Eprint [0]{\href }%
\providecommand \doibase [0]{https://doi.org/}%
\providecommand \selectlanguage [0]{\@gobble}%
\providecommand \bibinfo  [0]{\@secondoftwo}%
\providecommand \bibfield  [0]{\@secondoftwo}%
\providecommand \translation [1]{[#1]}%
\providecommand \BibitemOpen [0]{}%
\providecommand \bibitemStop [0]{}%
\providecommand \bibitemNoStop [0]{.\EOS\space}%
\providecommand \EOS [0]{\spacefactor3000\relax}%
\providecommand \BibitemShut  [1]{\csname bibitem#1\endcsname}%
\let\auto@bib@innerbib\@empty
\bibitem [{\citenamefont {Wilson}\ \emph {et~al.}(2021)\citenamefont {Wilson}, \citenamefont {Yao}, \citenamefont {Shan},\ and\ \citenamefont {Xu}}]{wilson2021excitons}%
  \BibitemOpen
  \bibfield  {author} {\bibinfo {author} {\bibfnamefont {N.~P.}\ \bibnamefont {Wilson}}, \bibinfo {author} {\bibfnamefont {W.}~\bibnamefont {Yao}}, \bibinfo {author} {\bibfnamefont {J.}~\bibnamefont {Shan}},\ and\ \bibinfo {author} {\bibfnamefont {X.}~\bibnamefont {Xu}},\ }\bibfield  {title} {\bibinfo {title} {Excitons and emergent quantum phenomena in stacked 2d semiconductors},\ }\href {https://doi.org/10.1038/s41586-021-03979-1} {\bibfield  {journal} {\bibinfo  {journal} {Nature}\ }\textbf {\bibinfo {volume} {599}},\ \bibinfo {pages} {383} (\bibinfo {year} {2021})}\BibitemShut {NoStop}%
\bibitem [{\citenamefont {Novoselov}\ \emph {et~al.}(2005)\citenamefont {Novoselov}, \citenamefont {Jiang}, \citenamefont {Schedin}, \citenamefont {Booth}, \citenamefont {Khotkevich}, \citenamefont {Morozov},\ and\ \citenamefont {Geim}}]{novoselov2005two}%
  \BibitemOpen
  \bibfield  {author} {\bibinfo {author} {\bibfnamefont {K.~S.}\ \bibnamefont {Novoselov}}, \bibinfo {author} {\bibfnamefont {D.}~\bibnamefont {Jiang}}, \bibinfo {author} {\bibfnamefont {F.}~\bibnamefont {Schedin}}, \bibinfo {author} {\bibfnamefont {T.}~\bibnamefont {Booth}}, \bibinfo {author} {\bibfnamefont {V.}~\bibnamefont {Khotkevich}}, \bibinfo {author} {\bibfnamefont {S.}~\bibnamefont {Morozov}},\ and\ \bibinfo {author} {\bibfnamefont {A.~K.}\ \bibnamefont {Geim}},\ }\bibfield  {title} {\bibinfo {title} {Two-dimensional atomic crystals},\ }\href {https://doi.org/10.1073/pnas.0502848102} {\bibfield  {journal} {\bibinfo  {journal} {Proceedings of the National Academy of Sciences}\ }\textbf {\bibinfo {volume} {102}},\ \bibinfo {pages} {10451} (\bibinfo {year} {2005})}\BibitemShut {NoStop}%
\bibitem [{\citenamefont {Splendiani}\ \emph {et~al.}(2010)\citenamefont {Splendiani}, \citenamefont {Sun}, \citenamefont {Zhang}, \citenamefont {Li}, \citenamefont {Kim}, \citenamefont {Chim}, \citenamefont {Galli},\ and\ \citenamefont {Wang}}]{splendiani2010emerging}%
  \BibitemOpen
  \bibfield  {author} {\bibinfo {author} {\bibfnamefont {A.}~\bibnamefont {Splendiani}}, \bibinfo {author} {\bibfnamefont {L.}~\bibnamefont {Sun}}, \bibinfo {author} {\bibfnamefont {Y.}~\bibnamefont {Zhang}}, \bibinfo {author} {\bibfnamefont {T.}~\bibnamefont {Li}}, \bibinfo {author} {\bibfnamefont {J.}~\bibnamefont {Kim}}, \bibinfo {author} {\bibfnamefont {C.-Y.}\ \bibnamefont {Chim}}, \bibinfo {author} {\bibfnamefont {G.}~\bibnamefont {Galli}},\ and\ \bibinfo {author} {\bibfnamefont {F.}~\bibnamefont {Wang}},\ }\bibfield  {title} {\bibinfo {title} {Emerging photoluminescence in monolayer mos2},\ }\href {https://doi.org/10.1021/nl903868w} {\bibfield  {journal} {\bibinfo  {journal} {Nano letters}\ }\textbf {\bibinfo {volume} {10}},\ \bibinfo {pages} {1271} (\bibinfo {year} {2010})}\BibitemShut {NoStop}%
\bibitem [{\citenamefont {Lee}\ \emph {et~al.}(2012)\citenamefont {Lee}, \citenamefont {Zhang}, \citenamefont {Zhang}, \citenamefont {Chang}, \citenamefont {Lin}, \citenamefont {Chang}, \citenamefont {Yu}, \citenamefont {Wang}, \citenamefont {Chang}, \citenamefont {Li},\ and\ \citenamefont {Lin}}]{lee2012synthesis}%
  \BibitemOpen
  \bibfield  {author} {\bibinfo {author} {\bibfnamefont {Y.-H.}\ \bibnamefont {Lee}}, \bibinfo {author} {\bibfnamefont {X.-Q.}\ \bibnamefont {Zhang}}, \bibinfo {author} {\bibfnamefont {W.}~\bibnamefont {Zhang}}, \bibinfo {author} {\bibfnamefont {M.-T.}\ \bibnamefont {Chang}}, \bibinfo {author} {\bibfnamefont {C.-T.}\ \bibnamefont {Lin}}, \bibinfo {author} {\bibfnamefont {K.-D.}\ \bibnamefont {Chang}}, \bibinfo {author} {\bibfnamefont {Y.-C.}\ \bibnamefont {Yu}}, \bibinfo {author} {\bibfnamefont {J.~T.-W.}\ \bibnamefont {Wang}}, \bibinfo {author} {\bibfnamefont {C.-S.}\ \bibnamefont {Chang}}, \bibinfo {author} {\bibfnamefont {L.-J.}\ \bibnamefont {Li}},\ and\ \bibinfo {author} {\bibfnamefont {T.-W.}\ \bibnamefont {Lin}},\ }\bibfield  {title} {\bibinfo {title} {Synthesis of large-area mos2 atomic layers with chemical vapor deposition},\ }\href {https://doi.org/https://doi.org/10.1002/adma.201104798} {\bibfield  {journal} {\bibinfo  {journal} {Advanced Materials}\ }\textbf {\bibinfo {volume} {24}},\ \bibinfo
  {pages} {2320} (\bibinfo {year} {2012})},\ \Eprint {https://arxiv.org/abs/https://onlinelibrary.wiley.com/doi/pdf/10.1002/adma.201104798} {https://onlinelibrary.wiley.com/doi/pdf/10.1002/adma.201104798} \BibitemShut {NoStop}%
\bibitem [{\citenamefont {Sohier}\ \emph {et~al.}(2023)\citenamefont {Sohier}, \citenamefont {de~Melo}, \citenamefont {Zanolli},\ and\ \citenamefont {Verstraete}}]{sohier2023impact}%
  \BibitemOpen
  \bibfield  {author} {\bibinfo {author} {\bibfnamefont {T.}~\bibnamefont {Sohier}}, \bibinfo {author} {\bibfnamefont {P.~M.}\ \bibnamefont {de~Melo}}, \bibinfo {author} {\bibfnamefont {Z.}~\bibnamefont {Zanolli}},\ and\ \bibinfo {author} {\bibfnamefont {M.~J.}\ \bibnamefont {Verstraete}},\ }\bibfield  {title} {\bibinfo {title} {The impact of valley profile on the mobility and kerr rotation of transition metal dichalcogenides},\ }\href {https://doi.org/10.1088/2053-1583/acb21c} {\bibfield  {journal} {\bibinfo  {journal} {2D Materials}\ }\textbf {\bibinfo {volume} {10}},\ \bibinfo {pages} {025006} (\bibinfo {year} {2023})}\BibitemShut {NoStop}%
\bibitem [{\citenamefont {Tran}\ \emph {et~al.}(2019)\citenamefont {Tran}, \citenamefont {Moody}, \citenamefont {Wu}, \citenamefont {Lu}, \citenamefont {Choi}, \citenamefont {Kim}, \citenamefont {Rai}, \citenamefont {Sanchez}, \citenamefont {Quan}, \citenamefont {Singh} \emph {et~al.}}]{tran2019evidence}%
  \BibitemOpen
  \bibfield  {author} {\bibinfo {author} {\bibfnamefont {K.}~\bibnamefont {Tran}}, \bibinfo {author} {\bibfnamefont {G.}~\bibnamefont {Moody}}, \bibinfo {author} {\bibfnamefont {F.}~\bibnamefont {Wu}}, \bibinfo {author} {\bibfnamefont {X.}~\bibnamefont {Lu}}, \bibinfo {author} {\bibfnamefont {J.}~\bibnamefont {Choi}}, \bibinfo {author} {\bibfnamefont {K.}~\bibnamefont {Kim}}, \bibinfo {author} {\bibfnamefont {A.}~\bibnamefont {Rai}}, \bibinfo {author} {\bibfnamefont {D.~A.}\ \bibnamefont {Sanchez}}, \bibinfo {author} {\bibfnamefont {J.}~\bibnamefont {Quan}}, \bibinfo {author} {\bibfnamefont {A.}~\bibnamefont {Singh}}, \emph {et~al.},\ }\bibfield  {title} {\bibinfo {title} {Evidence for moir{\'e} excitons in van der waals heterostructures},\ }\href {https://doi.org/10.1038/s41586-019-0975-z} {\bibfield  {journal} {\bibinfo  {journal} {Nature}\ }\textbf {\bibinfo {volume} {567}},\ \bibinfo {pages} {71} (\bibinfo {year} {2019})}\BibitemShut {NoStop}%
\bibitem [{\citenamefont {Ma}\ \emph {et~al.}(2023{\natexlab{a}})\citenamefont {Ma}, \citenamefont {Zhang}, \citenamefont {Li}, \citenamefont {Geng},\ and\ \citenamefont {Hu}}]{rakib2022moire}%
  \BibitemOpen
  \bibfield  {author} {\bibinfo {author} {\bibfnamefont {W.}~\bibnamefont {Ma}}, \bibinfo {author} {\bibfnamefont {Q.}~\bibnamefont {Zhang}}, \bibinfo {author} {\bibfnamefont {L.}~\bibnamefont {Li}}, \bibinfo {author} {\bibfnamefont {D.}~\bibnamefont {Geng}},\ and\ \bibinfo {author} {\bibfnamefont {W.}~\bibnamefont {Hu}},\ }\bibfield  {title} {\bibinfo {title} {Small twist{,} big miracle—recent progress in the fabrication of twisted 2d materials},\ }\href {https://doi.org/10.1039/D3TC02660D} {\bibfield  {journal} {\bibinfo  {journal} {J. Mater. Chem. C}\ }\textbf {\bibinfo {volume} {11}},\ \bibinfo {pages} {15793} (\bibinfo {year} {2023}{\natexlab{a}})}\BibitemShut {NoStop}%
\bibitem [{\citenamefont {Ma}\ \emph {et~al.}(2023{\natexlab{b}})\citenamefont {Ma}, \citenamefont {Zhang}, \citenamefont {Li}, \citenamefont {Geng},\ and\ \citenamefont {Hu}}]{ma2023small}%
  \BibitemOpen
  \bibfield  {author} {\bibinfo {author} {\bibfnamefont {W.}~\bibnamefont {Ma}}, \bibinfo {author} {\bibfnamefont {Q.}~\bibnamefont {Zhang}}, \bibinfo {author} {\bibfnamefont {L.}~\bibnamefont {Li}}, \bibinfo {author} {\bibfnamefont {D.}~\bibnamefont {Geng}},\ and\ \bibinfo {author} {\bibfnamefont {W.}~\bibnamefont {Hu}},\ }\bibfield  {title} {\bibinfo {title} {Small twist{,} big miracle—recent progress in the fabrication of twisted 2d materials},\ }\href {https://doi.org/10.1039/D3TC02660D} {\bibfield  {journal} {\bibinfo  {journal} {J. Mater. Chem. C}\ }\textbf {\bibinfo {volume} {11}},\ \bibinfo {pages} {15793} (\bibinfo {year} {2023}{\natexlab{b}})}\BibitemShut {NoStop}%
\bibitem [{\citenamefont {Ribeiro-Palau}\ \emph {et~al.}(2018)\citenamefont {Ribeiro-Palau}, \citenamefont {Zhang}, \citenamefont {Watanabe}, \citenamefont {Taniguchi}, \citenamefont {Hone},\ and\ \citenamefont {Dean}}]{ribeiro2018twistable}%
  \BibitemOpen
  \bibfield  {author} {\bibinfo {author} {\bibfnamefont {R.}~\bibnamefont {Ribeiro-Palau}}, \bibinfo {author} {\bibfnamefont {C.}~\bibnamefont {Zhang}}, \bibinfo {author} {\bibfnamefont {K.}~\bibnamefont {Watanabe}}, \bibinfo {author} {\bibfnamefont {T.}~\bibnamefont {Taniguchi}}, \bibinfo {author} {\bibfnamefont {J.}~\bibnamefont {Hone}},\ and\ \bibinfo {author} {\bibfnamefont {C.~R.}\ \bibnamefont {Dean}},\ }\bibfield  {title} {\bibinfo {title} {Twistable electronics with dynamically rotatable heterostructures},\ }\href {https://doi.org/10.1126/science.aat6981} {\bibfield  {journal} {\bibinfo  {journal} {Science}\ }\textbf {\bibinfo {volume} {361}},\ \bibinfo {pages} {690} (\bibinfo {year} {2018})}\BibitemShut {NoStop}%
\bibitem [{\citenamefont {Wang}\ \emph {et~al.}(2018)\citenamefont {Wang}, \citenamefont {Chernikov}, \citenamefont {Glazov}, \citenamefont {Heinz}, \citenamefont {Marie}, \citenamefont {Amand},\ and\ \citenamefont {Urbaszek}}]{wang2018colloquium}%
  \BibitemOpen
  \bibfield  {author} {\bibinfo {author} {\bibfnamefont {G.}~\bibnamefont {Wang}}, \bibinfo {author} {\bibfnamefont {A.}~\bibnamefont {Chernikov}}, \bibinfo {author} {\bibfnamefont {M.~M.}\ \bibnamefont {Glazov}}, \bibinfo {author} {\bibfnamefont {T.~F.}\ \bibnamefont {Heinz}}, \bibinfo {author} {\bibfnamefont {X.}~\bibnamefont {Marie}}, \bibinfo {author} {\bibfnamefont {T.}~\bibnamefont {Amand}},\ and\ \bibinfo {author} {\bibfnamefont {B.}~\bibnamefont {Urbaszek}},\ }\bibfield  {title} {\bibinfo {title} {Colloquium: Excitons in atomically thin transition metal dichalcogenides},\ }\href {https://doi.org/10.1103/RevModPhys.90.021001} {\bibfield  {journal} {\bibinfo  {journal} {Reviews of Modern Physics}\ }\textbf {\bibinfo {volume} {90}},\ \bibinfo {pages} {021001} (\bibinfo {year} {2018})}\BibitemShut {NoStop}%
\bibitem [{\citenamefont {Ji}\ \emph {et~al.}(2017)\citenamefont {Ji}, \citenamefont {Hong}, \citenamefont {Zhang}, \citenamefont {Zhang}, \citenamefont {Huang}, \citenamefont {Cao}, \citenamefont {Qiao}, \citenamefont {Liu}, \citenamefont {Liang}, \citenamefont {Jin} \emph {et~al.}}]{ji2017robust}%
  \BibitemOpen
  \bibfield  {author} {\bibinfo {author} {\bibfnamefont {Z.}~\bibnamefont {Ji}}, \bibinfo {author} {\bibfnamefont {H.}~\bibnamefont {Hong}}, \bibinfo {author} {\bibfnamefont {J.}~\bibnamefont {Zhang}}, \bibinfo {author} {\bibfnamefont {Q.}~\bibnamefont {Zhang}}, \bibinfo {author} {\bibfnamefont {W.}~\bibnamefont {Huang}}, \bibinfo {author} {\bibfnamefont {T.}~\bibnamefont {Cao}}, \bibinfo {author} {\bibfnamefont {R.}~\bibnamefont {Qiao}}, \bibinfo {author} {\bibfnamefont {C.}~\bibnamefont {Liu}}, \bibinfo {author} {\bibfnamefont {J.}~\bibnamefont {Liang}}, \bibinfo {author} {\bibfnamefont {C.}~\bibnamefont {Jin}}, \emph {et~al.},\ }\bibfield  {title} {\bibinfo {title} {Robust stacking-independent ultrafast charge transfer in mos2/ws2 bilayers},\ }\href {https://doi.org/10.1021/acsnano.7b04541} {\bibfield  {journal} {\bibinfo  {journal} {ACS nano}\ }\textbf {\bibinfo {volume} {11}},\ \bibinfo {pages} {12020} (\bibinfo {year} {2017})}\BibitemShut {NoStop}%
\bibitem [{\citenamefont {Trovatello}\ \emph {et~al.}(2020)\citenamefont {Trovatello}, \citenamefont {Katsch}, \citenamefont {Borys}, \citenamefont {Selig}, \citenamefont {Yao}, \citenamefont {Borrego-Varillas}, \citenamefont {Scotognella}, \citenamefont {Kriegel}, \citenamefont {Yan}, \citenamefont {Zettl} \emph {et~al.}}]{trovatello2020ultrafast}%
  \BibitemOpen
  \bibfield  {author} {\bibinfo {author} {\bibfnamefont {C.}~\bibnamefont {Trovatello}}, \bibinfo {author} {\bibfnamefont {F.}~\bibnamefont {Katsch}}, \bibinfo {author} {\bibfnamefont {N.~J.}\ \bibnamefont {Borys}}, \bibinfo {author} {\bibfnamefont {M.}~\bibnamefont {Selig}}, \bibinfo {author} {\bibfnamefont {K.}~\bibnamefont {Yao}}, \bibinfo {author} {\bibfnamefont {R.}~\bibnamefont {Borrego-Varillas}}, \bibinfo {author} {\bibfnamefont {F.}~\bibnamefont {Scotognella}}, \bibinfo {author} {\bibfnamefont {I.}~\bibnamefont {Kriegel}}, \bibinfo {author} {\bibfnamefont {A.}~\bibnamefont {Yan}}, \bibinfo {author} {\bibfnamefont {A.}~\bibnamefont {Zettl}}, \emph {et~al.},\ }\bibfield  {title} {\bibinfo {title} {The ultrafast onset of exciton formation in 2d semiconductors},\ }\href {https://doi.org/10.1038/s41467-020-18835-5} {\bibfield  {journal} {\bibinfo  {journal} {Nature communications}\ }\textbf {\bibinfo {volume} {11}},\ \bibinfo {pages} {5277} (\bibinfo {year} {2020})}\BibitemShut {NoStop}%
\bibitem [{\citenamefont {Tsai}\ \emph {et~al.}(2022)\citenamefont {Tsai}, \citenamefont {Pan}, \citenamefont {Lin}, \citenamefont {Bansil},\ and\ \citenamefont {Yan}}]{tsai2022antisite}%
  \BibitemOpen
  \bibfield  {author} {\bibinfo {author} {\bibfnamefont {J.-Y.}\ \bibnamefont {Tsai}}, \bibinfo {author} {\bibfnamefont {J.}~\bibnamefont {Pan}}, \bibinfo {author} {\bibfnamefont {H.}~\bibnamefont {Lin}}, \bibinfo {author} {\bibfnamefont {A.}~\bibnamefont {Bansil}},\ and\ \bibinfo {author} {\bibfnamefont {Q.}~\bibnamefont {Yan}},\ }\bibfield  {title} {\bibinfo {title} {Antisite defect qubits in monolayer transition metal dichalcogenides},\ }\href {https://doi.org/10.1038/s41467-022-28133-x} {\bibfield  {journal} {\bibinfo  {journal} {Nature communications}\ }\textbf {\bibinfo {volume} {13}},\ \bibinfo {pages} {492} (\bibinfo {year} {2022})}\BibitemShut {NoStop}%
\bibitem [{\citenamefont {Jiang}\ \emph {et~al.}(2021)\citenamefont {Jiang}, \citenamefont {Chen}, \citenamefont {Zheng}, \citenamefont {Zheng},\ and\ \citenamefont {Pan}}]{jiang2021interlayer}%
  \BibitemOpen
  \bibfield  {author} {\bibinfo {author} {\bibfnamefont {Y.}~\bibnamefont {Jiang}}, \bibinfo {author} {\bibfnamefont {S.}~\bibnamefont {Chen}}, \bibinfo {author} {\bibfnamefont {W.}~\bibnamefont {Zheng}}, \bibinfo {author} {\bibfnamefont {B.}~\bibnamefont {Zheng}},\ and\ \bibinfo {author} {\bibfnamefont {A.}~\bibnamefont {Pan}},\ }\bibfield  {title} {\bibinfo {title} {Interlayer exciton formation, relaxation, and transport in tmd van der waals heterostructures},\ }\href {https://doi.org/10.1038/s41377-021-00500-1} {\bibfield  {journal} {\bibinfo  {journal} {Light: Science \& Applications}\ }\textbf {\bibinfo {volume} {10}},\ \bibinfo {pages} {72} (\bibinfo {year} {2021})}\BibitemShut {NoStop}%
\bibitem [{\citenamefont {Torun}\ \emph {et~al.}(2018)\citenamefont {Torun}, \citenamefont {Miranda}, \citenamefont {Molina-S{\'a}nchez},\ and\ \citenamefont {Wirtz}}]{torun2018interlayer}%
  \BibitemOpen
  \bibfield  {author} {\bibinfo {author} {\bibfnamefont {E.}~\bibnamefont {Torun}}, \bibinfo {author} {\bibfnamefont {H.~P.}\ \bibnamefont {Miranda}}, \bibinfo {author} {\bibfnamefont {A.}~\bibnamefont {Molina-S{\'a}nchez}},\ and\ \bibinfo {author} {\bibfnamefont {L.}~\bibnamefont {Wirtz}},\ }\bibfield  {title} {\bibinfo {title} {Interlayer and intralayer excitons in mos2/ws2 and mose2/wse2 heterobilayers},\ }\href {https://doi.org/10.1103/PhysRevB.97.245427} {\bibfield  {journal} {\bibinfo  {journal} {Physical Review B}\ }\textbf {\bibinfo {volume} {97}},\ \bibinfo {pages} {245427} (\bibinfo {year} {2018})}\BibitemShut {NoStop}%
\bibitem [{\citenamefont {Gillen}\ and\ \citenamefont {Maultzsch}(2018)}]{gillen2018interlayer}%
  \BibitemOpen
  \bibfield  {author} {\bibinfo {author} {\bibfnamefont {R.}~\bibnamefont {Gillen}}\ and\ \bibinfo {author} {\bibfnamefont {J.}~\bibnamefont {Maultzsch}},\ }\bibfield  {title} {\bibinfo {title} {Interlayer excitons in mose2/wse2 heterostructures from first principles},\ }\href {https://doi.org/10.1103/Phys-RevB.97.165306} {\bibfield  {journal} {\bibinfo  {journal} {Physical Review B}\ }\textbf {\bibinfo {volume} {97}},\ \bibinfo {pages} {165306} (\bibinfo {year} {2018})}\BibitemShut {NoStop}%
\bibitem [{\citenamefont {Sangalli}\ \emph {et~al.}(2019)\citenamefont {Sangalli}, \citenamefont {Ferretti}, \citenamefont {Miranda}, \citenamefont {Attaccalite}, \citenamefont {Marri}, \citenamefont {Cannuccia}, \citenamefont {Melo}, \citenamefont {Marsili}, \citenamefont {Paleari}, \citenamefont {Marrazzo} \emph {et~al.}}]{sangalli2019many}%
  \BibitemOpen
  \bibfield  {author} {\bibinfo {author} {\bibfnamefont {D.}~\bibnamefont {Sangalli}}, \bibinfo {author} {\bibfnamefont {A.}~\bibnamefont {Ferretti}}, \bibinfo {author} {\bibfnamefont {H.}~\bibnamefont {Miranda}}, \bibinfo {author} {\bibfnamefont {C.}~\bibnamefont {Attaccalite}}, \bibinfo {author} {\bibfnamefont {I.}~\bibnamefont {Marri}}, \bibinfo {author} {\bibfnamefont {E.}~\bibnamefont {Cannuccia}}, \bibinfo {author} {\bibfnamefont {P.}~\bibnamefont {Melo}}, \bibinfo {author} {\bibfnamefont {M.}~\bibnamefont {Marsili}}, \bibinfo {author} {\bibfnamefont {F.}~\bibnamefont {Paleari}}, \bibinfo {author} {\bibfnamefont {A.}~\bibnamefont {Marrazzo}}, \emph {et~al.},\ }\bibfield  {title} {\bibinfo {title} {Many-body perturbation theory calculations using the yambo code},\ }\href {https://doi.org/10.1088/1361-648X/ab15d0} {\bibfield  {journal} {\bibinfo  {journal} {Journal of Physics: Condensed Matter}\ }\textbf {\bibinfo {volume} {31}},\ \bibinfo {pages} {325902} (\bibinfo {year} {2019})}\BibitemShut {NoStop}%
\bibitem [{\citenamefont {Marini}\ \emph {et~al.}(2009)\citenamefont {Marini}, \citenamefont {Hogan}, \citenamefont {Gr{\"u}ning},\ and\ \citenamefont {Varsano}}]{marini2009yambo}%
  \BibitemOpen
  \bibfield  {author} {\bibinfo {author} {\bibfnamefont {A.}~\bibnamefont {Marini}}, \bibinfo {author} {\bibfnamefont {C.}~\bibnamefont {Hogan}}, \bibinfo {author} {\bibfnamefont {M.}~\bibnamefont {Gr{\"u}ning}},\ and\ \bibinfo {author} {\bibfnamefont {D.}~\bibnamefont {Varsano}},\ }\bibfield  {title} {\bibinfo {title} {Yambo: An ab initio tool for excited state calculations},\ }\href {https://doi.org/10.1016/j.cpc.2009.02.003} {\bibfield  {journal} {\bibinfo  {journal} {Computer Physics Communications}\ }\textbf {\bibinfo {volume} {180}},\ \bibinfo {pages} {1392} (\bibinfo {year} {2009})}\BibitemShut {NoStop}%
\bibitem [{\citenamefont {de~Melo}\ and\ \citenamefont {Marini}(2016)}]{de2016unified}%
  \BibitemOpen
  \bibfield  {author} {\bibinfo {author} {\bibfnamefont {P.~M. M.~C.}\ \bibnamefont {de~Melo}}\ and\ \bibinfo {author} {\bibfnamefont {A.}~\bibnamefont {Marini}},\ }\bibfield  {title} {\bibinfo {title} {Unified theory of quantized electrons, phonons, and photons out of equilibrium: A simplified ab initio approach based on the generalized baym-kadanoff ansatz},\ }\href {https://doi.org/10.1103/PhysRevB.93.155102} {\bibfield  {journal} {\bibinfo  {journal} {Phys. Rev. B}\ }\textbf {\bibinfo {volume} {93}},\ \bibinfo {pages} {155102} (\bibinfo {year} {2016})}\BibitemShut {NoStop}%
\bibitem [{\citenamefont {Lechifflart}\ \emph {et~al.}(2023)\citenamefont {Lechifflart}, \citenamefont {Paleari}, \citenamefont {Sangalli},\ and\ \citenamefont {Attaccalite}}]{lechifflart2023first}%
  \BibitemOpen
  \bibfield  {author} {\bibinfo {author} {\bibfnamefont {P.}~\bibnamefont {Lechifflart}}, \bibinfo {author} {\bibfnamefont {F.}~\bibnamefont {Paleari}}, \bibinfo {author} {\bibfnamefont {D.}~\bibnamefont {Sangalli}},\ and\ \bibinfo {author} {\bibfnamefont {C.}~\bibnamefont {Attaccalite}},\ }\bibfield  {title} {\bibinfo {title} {First-principles study of luminescence in hexagonal boron nitride single layer: Exciton-phonon coupling and the role of substrate},\ }\href {https://doi.org/10.1103/PhysRevMateri- als.7.024006} {\bibfield  {journal} {\bibinfo  {journal} {Physical Review Materials}\ }\textbf {\bibinfo {volume} {7}},\ \bibinfo {pages} {024006} (\bibinfo {year} {2023})}\BibitemShut {NoStop}%
\bibitem [{\citenamefont {Paleari}\ \emph {et~al.}(2019)\citenamefont {Paleari}, \citenamefont {Miranda}, \citenamefont {Molina-S{\'a}nchez},\ and\ \citenamefont {Wirtz}}]{paleari2019exciton}%
  \BibitemOpen
  \bibfield  {author} {\bibinfo {author} {\bibfnamefont {F.}~\bibnamefont {Paleari}}, \bibinfo {author} {\bibfnamefont {H.~P.}\ \bibnamefont {Miranda}}, \bibinfo {author} {\bibfnamefont {A.}~\bibnamefont {Molina-S{\'a}nchez}},\ and\ \bibinfo {author} {\bibfnamefont {L.}~\bibnamefont {Wirtz}},\ }\bibfield  {title} {\bibinfo {title} {Exciton-phonon coupling in the ultraviolet absorption and emission spectra of bulk hexagonal boron nitride},\ }\href {https://doi.org/10.1103/PhysRevLett.122.187401} {\bibfield  {journal} {\bibinfo  {journal} {Physical review letters}\ }\textbf {\bibinfo {volume} {122}},\ \bibinfo {pages} {187401} (\bibinfo {year} {2019})}\BibitemShut {NoStop}%
\bibitem [{\citenamefont {Palummo}\ \emph {et~al.}(2015)\citenamefont {Palummo}, \citenamefont {Bernardi},\ and\ \citenamefont {Grossman}}]{palummo2015exciton}%
  \BibitemOpen
  \bibfield  {author} {\bibinfo {author} {\bibfnamefont {M.}~\bibnamefont {Palummo}}, \bibinfo {author} {\bibfnamefont {M.}~\bibnamefont {Bernardi}},\ and\ \bibinfo {author} {\bibfnamefont {J.~C.}\ \bibnamefont {Grossman}},\ }\bibfield  {title} {\bibinfo {title} {Exciton radiative lifetimes in two-dimensional transition metal dichalcogenides},\ }\href {https://doi.org/10.1021/nl503799t} {\bibfield  {journal} {\bibinfo  {journal} {Nano letters}\ }\textbf {\bibinfo {volume} {15}},\ \bibinfo {pages} {2794} (\bibinfo {year} {2015})}\BibitemShut {NoStop}%
\bibitem [{\citenamefont {Chen}\ \emph {et~al.}(2018)\citenamefont {Chen}, \citenamefont {Palummo}, \citenamefont {Sangalli},\ and\ \citenamefont {Bernardi}}]{chen2018theory}%
  \BibitemOpen
  \bibfield  {author} {\bibinfo {author} {\bibfnamefont {H.-Y.}\ \bibnamefont {Chen}}, \bibinfo {author} {\bibfnamefont {M.}~\bibnamefont {Palummo}}, \bibinfo {author} {\bibfnamefont {D.}~\bibnamefont {Sangalli}},\ and\ \bibinfo {author} {\bibfnamefont {M.}~\bibnamefont {Bernardi}},\ }\bibfield  {title} {\bibinfo {title} {Theory and ab initio computation of the anisotropic light emission in monolayer transition metal dichalcogenides},\ }\href {https://doi.org/10.1021/acs.nanolett.8b01114} {\bibfield  {journal} {\bibinfo  {journal} {Nano letters}\ }\textbf {\bibinfo {volume} {18}},\ \bibinfo {pages} {3839} (\bibinfo {year} {2018})}\BibitemShut {NoStop}%
\bibitem [{\citenamefont {Giannozzi}\ \emph {et~al.}(2009)\citenamefont {Giannozzi}, \citenamefont {Baroni}, \citenamefont {Bonini}, \citenamefont {Calandra}, \citenamefont {Car}, \citenamefont {Cavazzoni}, \citenamefont {Ceresoli}, \citenamefont {Chiarotti}, \citenamefont {Cococcioni}, \citenamefont {Dabo} \emph {et~al.}}]{giannozzi2009quantum}%
  \BibitemOpen
  \bibfield  {author} {\bibinfo {author} {\bibfnamefont {P.}~\bibnamefont {Giannozzi}}, \bibinfo {author} {\bibfnamefont {S.}~\bibnamefont {Baroni}}, \bibinfo {author} {\bibfnamefont {N.}~\bibnamefont {Bonini}}, \bibinfo {author} {\bibfnamefont {M.}~\bibnamefont {Calandra}}, \bibinfo {author} {\bibfnamefont {R.}~\bibnamefont {Car}}, \bibinfo {author} {\bibfnamefont {C.}~\bibnamefont {Cavazzoni}}, \bibinfo {author} {\bibfnamefont {D.}~\bibnamefont {Ceresoli}}, \bibinfo {author} {\bibfnamefont {G.~L.}\ \bibnamefont {Chiarotti}}, \bibinfo {author} {\bibfnamefont {M.}~\bibnamefont {Cococcioni}}, \bibinfo {author} {\bibfnamefont {I.}~\bibnamefont {Dabo}}, \emph {et~al.},\ }\bibfield  {title} {\bibinfo {title} {Quantum espresso: a modular and open-source software project for quantum simulations of materials},\ }\href {https://doi.org/10.1088/0953-8984/21/39/395502} {\bibfield  {journal} {\bibinfo  {journal} {Journal of physics: Condensed matter}\ }\textbf {\bibinfo {volume} {21}},\ \bibinfo {pages} {395502}
  (\bibinfo {year} {2009})}\BibitemShut {NoStop}%
\bibitem [{\citenamefont {Giannozzi}\ \emph {et~al.}(2017)\citenamefont {Giannozzi}, \citenamefont {Andreussi}, \citenamefont {Brumme}, \citenamefont {Bunau}, \citenamefont {Nardelli}, \citenamefont {Calandra}, \citenamefont {Car}, \citenamefont {Cavazzoni}, \citenamefont {Ceresoli}, \citenamefont {Cococcioni} \emph {et~al.}}]{giannozzi2017advanced}%
  \BibitemOpen
  \bibfield  {author} {\bibinfo {author} {\bibfnamefont {P.}~\bibnamefont {Giannozzi}}, \bibinfo {author} {\bibfnamefont {O.}~\bibnamefont {Andreussi}}, \bibinfo {author} {\bibfnamefont {T.}~\bibnamefont {Brumme}}, \bibinfo {author} {\bibfnamefont {O.}~\bibnamefont {Bunau}}, \bibinfo {author} {\bibfnamefont {M.~B.}\ \bibnamefont {Nardelli}}, \bibinfo {author} {\bibfnamefont {M.}~\bibnamefont {Calandra}}, \bibinfo {author} {\bibfnamefont {R.}~\bibnamefont {Car}}, \bibinfo {author} {\bibfnamefont {C.}~\bibnamefont {Cavazzoni}}, \bibinfo {author} {\bibfnamefont {D.}~\bibnamefont {Ceresoli}}, \bibinfo {author} {\bibfnamefont {M.}~\bibnamefont {Cococcioni}}, \emph {et~al.},\ }\bibfield  {title} {\bibinfo {title} {Advanced capabilities for materials modelling with quantum espresso},\ }\href {https://doi.org/10.1088/1361-648X/aa8f79} {\bibfield  {journal} {\bibinfo  {journal} {Journal of physics: Condensed matter}\ }\textbf {\bibinfo {volume} {29}},\ \bibinfo {pages} {465901} (\bibinfo {year} {2017})}\BibitemShut
  {NoStop}%
\bibitem [{\citenamefont {Barone}\ \emph {et~al.}(2009)\citenamefont {Barone}, \citenamefont {Casarin}, \citenamefont {Forrer}, \citenamefont {Pavone}, \citenamefont {Sambi},\ and\ \citenamefont {Vittadini}}]{barone2009role}%
  \BibitemOpen
  \bibfield  {author} {\bibinfo {author} {\bibfnamefont {V.}~\bibnamefont {Barone}}, \bibinfo {author} {\bibfnamefont {M.}~\bibnamefont {Casarin}}, \bibinfo {author} {\bibfnamefont {D.}~\bibnamefont {Forrer}}, \bibinfo {author} {\bibfnamefont {M.}~\bibnamefont {Pavone}}, \bibinfo {author} {\bibfnamefont {M.}~\bibnamefont {Sambi}},\ and\ \bibinfo {author} {\bibfnamefont {A.}~\bibnamefont {Vittadini}},\ }\bibfield  {title} {\bibinfo {title} {Role and effective treatment of dispersive forces in materials: Polyethylene and graphite crystals as test cases},\ }\href {https://doi.org/10.1002/jcc.21112} {\bibfield  {journal} {\bibinfo  {journal} {Journal of computational chemistry}\ }\textbf {\bibinfo {volume} {30}},\ \bibinfo {pages} {934} (\bibinfo {year} {2009})}\BibitemShut {NoStop}%
\bibitem [{\citenamefont {Hedin}(1965)}]{hedin1965new}%
  \BibitemOpen
  \bibfield  {author} {\bibinfo {author} {\bibfnamefont {L.}~\bibnamefont {Hedin}},\ }\bibfield  {title} {\bibinfo {title} {New method for calculating the one-particle green's function with application to the electron-gas problem},\ }\href {https://doi.org/10.1103/Phys- Rev.139.A796} {\bibfield  {journal} {\bibinfo  {journal} {Physical Review}\ }\textbf {\bibinfo {volume} {139}},\ \bibinfo {pages} {A796} (\bibinfo {year} {1965})}\BibitemShut {NoStop}%
\bibitem [{\citenamefont {Farid}\ \emph {et~al.}(1988)\citenamefont {Farid}, \citenamefont {Daling}, \citenamefont {Lenstra},\ and\ \citenamefont {van Haeringen}}]{farid1988gw}%
  \BibitemOpen
  \bibfield  {author} {\bibinfo {author} {\bibfnamefont {B.}~\bibnamefont {Farid}}, \bibinfo {author} {\bibfnamefont {R.}~\bibnamefont {Daling}}, \bibinfo {author} {\bibfnamefont {D.}~\bibnamefont {Lenstra}},\ and\ \bibinfo {author} {\bibfnamefont {W.}~\bibnamefont {van Haeringen}},\ }\bibfield  {title} {\bibinfo {title} {Gw approach to the calculation of electron self-energies in semiconductors},\ }\href {https://doi.org/10.1103/PhysRevB.38.7530} {\bibfield  {journal} {\bibinfo  {journal} {Physical Review B}\ }\textbf {\bibinfo {volume} {38}},\ \bibinfo {pages} {7530} (\bibinfo {year} {1988})}\BibitemShut {NoStop}%
\bibitem [{\citenamefont {Guandalini}\ \emph {et~al.}(2023)\citenamefont {Guandalini}, \citenamefont {D’Amico}, \citenamefont {Ferretti},\ and\ \citenamefont {Varsano}}]{guandalini2023efficient}%
  \BibitemOpen
  \bibfield  {author} {\bibinfo {author} {\bibfnamefont {A.}~\bibnamefont {Guandalini}}, \bibinfo {author} {\bibfnamefont {P.}~\bibnamefont {D’Amico}}, \bibinfo {author} {\bibfnamefont {A.}~\bibnamefont {Ferretti}},\ and\ \bibinfo {author} {\bibfnamefont {D.}~\bibnamefont {Varsano}},\ }\bibfield  {title} {\bibinfo {title} {Efficient gw calculations in two dimensional materials through a stochastic integration of the screened potential},\ }\href {https://doi.org/10.1038/s41524-023-00989-7} {\bibfield  {journal} {\bibinfo  {journal} {npj Computational Materials}\ }\textbf {\bibinfo {volume} {9}},\ \bibinfo {pages} {44} (\bibinfo {year} {2023})}\BibitemShut {NoStop}%
\bibitem [{\citenamefont {Onida}\ \emph {et~al.}(2002)\citenamefont {Onida}, \citenamefont {Reining},\ and\ \citenamefont {Rubio}}]{onida2002electronic}%
  \BibitemOpen
  \bibfield  {author} {\bibinfo {author} {\bibfnamefont {G.}~\bibnamefont {Onida}}, \bibinfo {author} {\bibfnamefont {L.}~\bibnamefont {Reining}},\ and\ \bibinfo {author} {\bibfnamefont {A.}~\bibnamefont {Rubio}},\ }\bibfield  {title} {\bibinfo {title} {Electronic excitations: density-functional versus many-body green’s-function approaches},\ }\href {https://doi.org/10.1103/RevModPhys.74.601} {\bibfield  {journal} {\bibinfo  {journal} {Reviews of modern physics}\ }\textbf {\bibinfo {volume} {74}},\ \bibinfo {pages} {601} (\bibinfo {year} {2002})}\BibitemShut {NoStop}%
\bibitem [{\citenamefont {Sangalli}\ \emph {et~al.}(2017)\citenamefont {Sangalli}, \citenamefont {Berger}, \citenamefont {Attaccalite}, \citenamefont {Gr{\"u}ning},\ and\ \citenamefont {Romaniello}}]{sangalli2017optical}%
  \BibitemOpen
  \bibfield  {author} {\bibinfo {author} {\bibfnamefont {D.}~\bibnamefont {Sangalli}}, \bibinfo {author} {\bibfnamefont {J.}~\bibnamefont {Berger}}, \bibinfo {author} {\bibfnamefont {C.}~\bibnamefont {Attaccalite}}, \bibinfo {author} {\bibfnamefont {M.}~\bibnamefont {Gr{\"u}ning}},\ and\ \bibinfo {author} {\bibfnamefont {P.}~\bibnamefont {Romaniello}},\ }\bibfield  {title} {\bibinfo {title} {Optical properties of periodic systems within the current-current response framework: Pitfalls and remedies},\ }\href {https://doi.org/10.1103/PhysRevB.95.155203} {\bibfield  {journal} {\bibinfo  {journal} {Physical Review B}\ }\textbf {\bibinfo {volume} {95}},\ \bibinfo {pages} {155203} (\bibinfo {year} {2017})}\BibitemShut {NoStop}%
\bibitem [{\citenamefont {Libbi}\ \emph {et~al.}(2022)\citenamefont {Libbi}, \citenamefont {de~Melo}, \citenamefont {Zanolli}, \citenamefont {Verstraete},\ and\ \citenamefont {Marzari}}]{libbi2022phonon}%
  \BibitemOpen
  \bibfield  {author} {\bibinfo {author} {\bibfnamefont {F.}~\bibnamefont {Libbi}}, \bibinfo {author} {\bibfnamefont {P.~M.~M.}\ \bibnamefont {de~Melo}}, \bibinfo {author} {\bibfnamefont {Z.}~\bibnamefont {Zanolli}}, \bibinfo {author} {\bibfnamefont {M.~J.}\ \bibnamefont {Verstraete}},\ and\ \bibinfo {author} {\bibfnamefont {N.}~\bibnamefont {Marzari}},\ }\bibfield  {title} {\bibinfo {title} {Phonon-assisted luminescence in defect centers from many-body perturbation theory},\ }\href {https://doi.org/10.1103/PhysRevLett.128.167401} {\bibfield  {journal} {\bibinfo  {journal} {Physical Review Letters}\ }\textbf {\bibinfo {volume} {128}},\ \bibinfo {pages} {167401} (\bibinfo {year} {2022})}\BibitemShut {NoStop}%
\bibitem [{\citenamefont {Hong}\ \emph {et~al.}(2014)\citenamefont {Hong}, \citenamefont {Kim}, \citenamefont {Shi}, \citenamefont {Zhang}, \citenamefont {Jin}, \citenamefont {Sun}, \citenamefont {Tongay}, \citenamefont {Wu}, \citenamefont {Zhang},\ and\ \citenamefont {Wang}}]{hong2014ultrafast}%
  \BibitemOpen
  \bibfield  {author} {\bibinfo {author} {\bibfnamefont {X.}~\bibnamefont {Hong}}, \bibinfo {author} {\bibfnamefont {J.}~\bibnamefont {Kim}}, \bibinfo {author} {\bibfnamefont {S.-F.}\ \bibnamefont {Shi}}, \bibinfo {author} {\bibfnamefont {Y.}~\bibnamefont {Zhang}}, \bibinfo {author} {\bibfnamefont {C.}~\bibnamefont {Jin}}, \bibinfo {author} {\bibfnamefont {Y.}~\bibnamefont {Sun}}, \bibinfo {author} {\bibfnamefont {S.}~\bibnamefont {Tongay}}, \bibinfo {author} {\bibfnamefont {J.}~\bibnamefont {Wu}}, \bibinfo {author} {\bibfnamefont {Y.}~\bibnamefont {Zhang}},\ and\ \bibinfo {author} {\bibfnamefont {F.}~\bibnamefont {Wang}},\ }\bibfield  {title} {\bibinfo {title} {Ultrafast charge transfer in atomically thin mos2/ws2 heterostructures},\ }\href {https://doi.org/10.1038/nnano.2014.167} {\bibfield  {journal} {\bibinfo  {journal} {Nature nanotechnology}\ }\textbf {\bibinfo {volume} {9}},\ \bibinfo {pages} {682} (\bibinfo {year} {2014})}\BibitemShut {NoStop}%
\bibitem [{\citenamefont {Heo}\ \emph {et~al.}(2015)\citenamefont {Heo}, \citenamefont {Sung}, \citenamefont {Cha}, \citenamefont {Jang}, \citenamefont {Kim}, \citenamefont {Jin}, \citenamefont {Lee}, \citenamefont {Ahn}, \citenamefont {Lee}, \citenamefont {Shim} \emph {et~al.}}]{heo2015interlayer}%
  \BibitemOpen
  \bibfield  {author} {\bibinfo {author} {\bibfnamefont {H.}~\bibnamefont {Heo}}, \bibinfo {author} {\bibfnamefont {J.~H.}\ \bibnamefont {Sung}}, \bibinfo {author} {\bibfnamefont {S.}~\bibnamefont {Cha}}, \bibinfo {author} {\bibfnamefont {B.-G.}\ \bibnamefont {Jang}}, \bibinfo {author} {\bibfnamefont {J.-Y.}\ \bibnamefont {Kim}}, \bibinfo {author} {\bibfnamefont {G.}~\bibnamefont {Jin}}, \bibinfo {author} {\bibfnamefont {D.}~\bibnamefont {Lee}}, \bibinfo {author} {\bibfnamefont {J.-H.}\ \bibnamefont {Ahn}}, \bibinfo {author} {\bibfnamefont {M.-J.}\ \bibnamefont {Lee}}, \bibinfo {author} {\bibfnamefont {J.~H.}\ \bibnamefont {Shim}}, \emph {et~al.},\ }\bibfield  {title} {\bibinfo {title} {Interlayer orientation-dependent light absorption and emission in monolayer semiconductor stacks},\ }\href {https://doi.org/10.1038/ncomms8372} {\bibfield  {journal} {\bibinfo  {journal} {Nature communications}\ }\textbf {\bibinfo {volume} {6}},\ \bibinfo {pages} {7372} (\bibinfo {year} {2015})}\BibitemShut {NoStop}%
\bibitem [{\citenamefont {Chen}\ \emph {et~al.}(2016)\citenamefont {Chen}, \citenamefont {Wen}, \citenamefont {Zhang}, \citenamefont {Wu}, \citenamefont {Gong}, \citenamefont {Zhang}, \citenamefont {Yuan}, \citenamefont {Yi}, \citenamefont {Lou}, \citenamefont {Ajayan} \emph {et~al.}}]{chen2016ultrafast}%
  \BibitemOpen
  \bibfield  {author} {\bibinfo {author} {\bibfnamefont {H.}~\bibnamefont {Chen}}, \bibinfo {author} {\bibfnamefont {X.}~\bibnamefont {Wen}}, \bibinfo {author} {\bibfnamefont {J.}~\bibnamefont {Zhang}}, \bibinfo {author} {\bibfnamefont {T.}~\bibnamefont {Wu}}, \bibinfo {author} {\bibfnamefont {Y.}~\bibnamefont {Gong}}, \bibinfo {author} {\bibfnamefont {X.}~\bibnamefont {Zhang}}, \bibinfo {author} {\bibfnamefont {J.}~\bibnamefont {Yuan}}, \bibinfo {author} {\bibfnamefont {C.}~\bibnamefont {Yi}}, \bibinfo {author} {\bibfnamefont {J.}~\bibnamefont {Lou}}, \bibinfo {author} {\bibfnamefont {P.~M.}\ \bibnamefont {Ajayan}}, \emph {et~al.},\ }\bibfield  {title} {\bibinfo {title} {Ultrafast formation of interlayer hot excitons in atomically thin mos2/ws2 heterostructures},\ }\href {https://doi.org/10.1038/ncomms12512} {\bibfield  {journal} {\bibinfo  {journal} {Nature communications}\ }\textbf {\bibinfo {volume} {7}},\ \bibinfo {pages} {12512} (\bibinfo {year} {2016})}\BibitemShut {NoStop}%
\bibitem [{\citenamefont {Li}\ \emph {et~al.}(2023)\citenamefont {Li}, \citenamefont {Gillen}, \citenamefont {Palummo}, \citenamefont {Milošević},\ and\ \citenamefont {Peeters}}]{li2023strain}%
  \BibitemOpen
  \bibfield  {author} {\bibinfo {author} {\bibfnamefont {L.~L.}\ \bibnamefont {Li}}, \bibinfo {author} {\bibfnamefont {R.}~\bibnamefont {Gillen}}, \bibinfo {author} {\bibfnamefont {M.}~\bibnamefont {Palummo}}, \bibinfo {author} {\bibfnamefont {M.~V.}\ \bibnamefont {Milošević}},\ and\ \bibinfo {author} {\bibfnamefont {F.~M.}\ \bibnamefont {Peeters}},\ }\bibfield  {title} {\bibinfo {title} {{Strain tunable interlayer and intralayer excitons in vertically stacked MoSe2/WSe2 heterobilayers}},\ }\href {https://doi.org/10.1063/5.0147761} {\bibfield  {journal} {\bibinfo  {journal} {Applied Physics Letters}\ }\textbf {\bibinfo {volume} {123}},\ \bibinfo {pages} {033102} (\bibinfo {year} {2023})}\BibitemShut {NoStop}%
\bibitem [{Note1()}]{Note1}%
  \BibitemOpen
  \bibinfo {note} {Https://github.com/rreho/yambopy}\BibitemShut {NoStop}%
\bibitem [{\citenamefont {{van Setten}}\ \emph {et~al.}(2018)\citenamefont {{van Setten}}, \citenamefont {Giantomassi}, \citenamefont {Bousquet}, \citenamefont {Verstraete}, \citenamefont {Hamann}, \citenamefont {Gonze},\ and\ \citenamefont {Rignanese}}]{vanSettenPseudoDojoTraining2018}%
  \BibitemOpen
  \bibfield  {author} {\bibinfo {author} {\bibfnamefont {M.}~\bibnamefont {{van Setten}}}, \bibinfo {author} {\bibfnamefont {M.}~\bibnamefont {Giantomassi}}, \bibinfo {author} {\bibfnamefont {E.}~\bibnamefont {Bousquet}}, \bibinfo {author} {\bibfnamefont {M.}~\bibnamefont {Verstraete}}, \bibinfo {author} {\bibfnamefont {D.}~\bibnamefont {Hamann}}, \bibinfo {author} {\bibfnamefont {X.}~\bibnamefont {Gonze}},\ and\ \bibinfo {author} {\bibfnamefont {G.-M.}\ \bibnamefont {Rignanese}},\ }\bibfield  {title} {\bibinfo {title} {The {{PseudoDojo}}: {{Training}} and grading a 85 element optimized norm-conserving pseudopotential table},\ }\href {https://doi.org/10.1016/j.cpc.2018.01.012} {\bibfield  {journal} {\bibinfo  {journal} {Computer Physics Communications}\ }\textbf {\bibinfo {volume} {226}},\ \bibinfo {pages} {39} (\bibinfo {year} {2018})}\BibitemShut {NoStop}%
\bibitem [{\citenamefont {Head}\ and\ \citenamefont {Zerner}(1985)}]{head1985broyden}%
  \BibitemOpen
  \bibfield  {author} {\bibinfo {author} {\bibfnamefont {J.~D.}\ \bibnamefont {Head}}\ and\ \bibinfo {author} {\bibfnamefont {M.~C.}\ \bibnamefont {Zerner}},\ }\bibfield  {title} {\bibinfo {title} {A broyden—fletcher—goldfarb—shanno optimization procedure for molecular geometries},\ }\href {https://doi.org/10.1016/0009-2614(85)80574-1} {\bibfield  {journal} {\bibinfo  {journal} {Chemical physics letters}\ }\textbf {\bibinfo {volume} {122}},\ \bibinfo {pages} {264} (\bibinfo {year} {1985})}\BibitemShut {NoStop}%
\bibitem [{\citenamefont {Grimme}\ \emph {et~al.}(2010)\citenamefont {Grimme}, \citenamefont {Antony}, \citenamefont {Ehrlich},\ and\ \citenamefont {Krieg}}]{grimme2010consistent}%
  \BibitemOpen
  \bibfield  {author} {\bibinfo {author} {\bibfnamefont {S.}~\bibnamefont {Grimme}}, \bibinfo {author} {\bibfnamefont {J.}~\bibnamefont {Antony}}, \bibinfo {author} {\bibfnamefont {S.}~\bibnamefont {Ehrlich}},\ and\ \bibinfo {author} {\bibfnamefont {H.}~\bibnamefont {Krieg}},\ }\bibfield  {title} {\bibinfo {title} {A consistent and accurate ab initio parametrization of density functional dispersion correction (dft-d) for the 94 elements h-pu},\ }\bibfield  {journal} {\bibinfo  {journal} {The Journal of chemical physics}\ }\textbf {\bibinfo {volume} {132}},\ \href {https://doi.org/10.1063/1.3382344} {10.1063/1.3382344} (\bibinfo {year} {2010})\BibitemShut {NoStop}%
\bibitem [{\citenamefont {Perdew}\ \emph {et~al.}(1996)\citenamefont {Perdew}, \citenamefont {Burke},\ and\ \citenamefont {Ernzerhof}}]{PerdewGeneralizedGradient1996}%
  \BibitemOpen
  \bibfield  {author} {\bibinfo {author} {\bibfnamefont {J.~P.}\ \bibnamefont {Perdew}}, \bibinfo {author} {\bibfnamefont {K.}~\bibnamefont {Burke}},\ and\ \bibinfo {author} {\bibfnamefont {M.}~\bibnamefont {Ernzerhof}},\ }\bibfield  {title} {\bibinfo {title} {Generalized {{Gradient Approximation Made Simple}}},\ }\href {https://doi.org/10.1103/PhysRevLett.77.3865} {\bibfield  {journal} {\bibinfo  {journal} {Physical Review Letters}\ }\textbf {\bibinfo {volume} {77}},\ \bibinfo {pages} {3865} (\bibinfo {year} {1996})}\BibitemShut {NoStop}%
\bibitem [{\citenamefont {Mak}\ \emph {et~al.}(2010)\citenamefont {Mak}, \citenamefont {Lee}, \citenamefont {Hone}, \citenamefont {Shan},\ and\ \citenamefont {Heinz}}]{mak2010atomically}%
  \BibitemOpen
  \bibfield  {author} {\bibinfo {author} {\bibfnamefont {K.~F.}\ \bibnamefont {Mak}}, \bibinfo {author} {\bibfnamefont {C.}~\bibnamefont {Lee}}, \bibinfo {author} {\bibfnamefont {J.}~\bibnamefont {Hone}}, \bibinfo {author} {\bibfnamefont {J.}~\bibnamefont {Shan}},\ and\ \bibinfo {author} {\bibfnamefont {T.~F.}\ \bibnamefont {Heinz}},\ }\bibfield  {title} {\bibinfo {title} {Atomically thin mos 2: a new direct-gap semiconductor},\ }\href {https://doi.org/10.1103/Phys- RevLett.105.136805} {\bibfield  {journal} {\bibinfo  {journal} {Physical review letters}\ }\textbf {\bibinfo {volume} {105}},\ \bibinfo {pages} {136805} (\bibinfo {year} {2010})}\BibitemShut {NoStop}%
\bibitem [{\citenamefont {Chhowalla}\ \emph {et~al.}(2013)\citenamefont {Chhowalla}, \citenamefont {Shin}, \citenamefont {Eda}, \citenamefont {Li}, \citenamefont {Loh},\ and\ \citenamefont {Zhang}}]{chhowalla2013chemistry}%
  \BibitemOpen
  \bibfield  {author} {\bibinfo {author} {\bibfnamefont {M.}~\bibnamefont {Chhowalla}}, \bibinfo {author} {\bibfnamefont {H.~S.}\ \bibnamefont {Shin}}, \bibinfo {author} {\bibfnamefont {G.}~\bibnamefont {Eda}}, \bibinfo {author} {\bibfnamefont {L.-J.}\ \bibnamefont {Li}}, \bibinfo {author} {\bibfnamefont {K.~P.}\ \bibnamefont {Loh}},\ and\ \bibinfo {author} {\bibfnamefont {H.}~\bibnamefont {Zhang}},\ }\bibfield  {title} {\bibinfo {title} {The chemistry of two-dimensional layered transition metal dichalcogenide nanosheets},\ }\href {https://doi.org/10.1038/nchem.1589} {\bibfield  {journal} {\bibinfo  {journal} {Nature chemistry}\ }\textbf {\bibinfo {volume} {5}},\ \bibinfo {pages} {263} (\bibinfo {year} {2013})}\BibitemShut {NoStop}%
\bibitem [{\citenamefont {Zhang}\ \emph {et~al.}(2014)\citenamefont {Zhang}, \citenamefont {Chang}, \citenamefont {Zhou}, \citenamefont {Cui}, \citenamefont {Yan}, \citenamefont {Liu}, \citenamefont {Schmitt}, \citenamefont {Lee}, \citenamefont {Moore}, \citenamefont {Chen} \emph {et~al.}}]{zhang2014direct}%
  \BibitemOpen
  \bibfield  {author} {\bibinfo {author} {\bibfnamefont {Y.}~\bibnamefont {Zhang}}, \bibinfo {author} {\bibfnamefont {T.-R.}\ \bibnamefont {Chang}}, \bibinfo {author} {\bibfnamefont {B.}~\bibnamefont {Zhou}}, \bibinfo {author} {\bibfnamefont {Y.-T.}\ \bibnamefont {Cui}}, \bibinfo {author} {\bibfnamefont {H.}~\bibnamefont {Yan}}, \bibinfo {author} {\bibfnamefont {Z.}~\bibnamefont {Liu}}, \bibinfo {author} {\bibfnamefont {F.}~\bibnamefont {Schmitt}}, \bibinfo {author} {\bibfnamefont {J.}~\bibnamefont {Lee}}, \bibinfo {author} {\bibfnamefont {R.}~\bibnamefont {Moore}}, \bibinfo {author} {\bibfnamefont {Y.}~\bibnamefont {Chen}}, \emph {et~al.},\ }\bibfield  {title} {\bibinfo {title} {Direct observation of the transition from indirect to direct bandgap in atomically thin epitaxial mose2},\ }\href {https://doi.org/10.1038/nnano.2013.277} {\bibfield  {journal} {\bibinfo  {journal} {Nature nanotechnology}\ }\textbf {\bibinfo {volume} {9}},\ \bibinfo {pages} {111} (\bibinfo {year} {2014})}\BibitemShut {NoStop}%
\bibitem [{\citenamefont {Zhao}\ \emph {et~al.}(2013)\citenamefont {Zhao}, \citenamefont {Ghorannevis}, \citenamefont {Chu}, \citenamefont {Toh}, \citenamefont {Kloc}, \citenamefont {Tan},\ and\ \citenamefont {Eda}}]{zhao2013evolution}%
  \BibitemOpen
  \bibfield  {author} {\bibinfo {author} {\bibfnamefont {W.}~\bibnamefont {Zhao}}, \bibinfo {author} {\bibfnamefont {Z.}~\bibnamefont {Ghorannevis}}, \bibinfo {author} {\bibfnamefont {L.}~\bibnamefont {Chu}}, \bibinfo {author} {\bibfnamefont {M.}~\bibnamefont {Toh}}, \bibinfo {author} {\bibfnamefont {C.}~\bibnamefont {Kloc}}, \bibinfo {author} {\bibfnamefont {P.-H.}\ \bibnamefont {Tan}},\ and\ \bibinfo {author} {\bibfnamefont {G.}~\bibnamefont {Eda}},\ }\bibfield  {title} {\bibinfo {title} {Evolution of electronic structure in atomically thin sheets of ws2 and wse2},\ }\href {https://doi.org/10.1021/nn305275h} {\bibfield  {journal} {\bibinfo  {journal} {ACS nano}\ }\textbf {\bibinfo {volume} {7}},\ \bibinfo {pages} {791} (\bibinfo {year} {2013})}\BibitemShut {NoStop}%
\bibitem [{\citenamefont {Lan}\ \emph {et~al.}(2018)\citenamefont {Lan}, \citenamefont {Yang}, \citenamefont {Xu}, \citenamefont {Qian}, \citenamefont {Zhang}, \citenamefont {Cheng},\ and\ \citenamefont {Zhang}}]{lan2018synthesis}%
  \BibitemOpen
  \bibfield  {author} {\bibinfo {author} {\bibfnamefont {F.}~\bibnamefont {Lan}}, \bibinfo {author} {\bibfnamefont {R.}~\bibnamefont {Yang}}, \bibinfo {author} {\bibfnamefont {Y.}~\bibnamefont {Xu}}, \bibinfo {author} {\bibfnamefont {S.}~\bibnamefont {Qian}}, \bibinfo {author} {\bibfnamefont {S.}~\bibnamefont {Zhang}}, \bibinfo {author} {\bibfnamefont {H.}~\bibnamefont {Cheng}},\ and\ \bibinfo {author} {\bibfnamefont {Y.}~\bibnamefont {Zhang}},\ }\bibfield  {title} {\bibinfo {title} {Synthesis of large-scale single-crystalline monolayer ws2 using a semi-sealed method},\ }\href {https://doi.org/10.3390/nano8020100} {\bibfield  {journal} {\bibinfo  {journal} {Nanomaterials}\ }\textbf {\bibinfo {volume} {8}},\ \bibinfo {pages} {100} (\bibinfo {year} {2018})}\BibitemShut {NoStop}%
\bibitem [{\citenamefont {Hanbicki}\ \emph {et~al.}(2018)\citenamefont {Hanbicki}, \citenamefont {Chuang}, \citenamefont {Rosenberger}, \citenamefont {Hellberg}, \citenamefont {Sivaram}, \citenamefont {McCreary}, \citenamefont {Mazin},\ and\ \citenamefont {Jonker}}]{hanbicki2018double}%
  \BibitemOpen
  \bibfield  {author} {\bibinfo {author} {\bibfnamefont {A.~T.}\ \bibnamefont {Hanbicki}}, \bibinfo {author} {\bibfnamefont {H.-J.}\ \bibnamefont {Chuang}}, \bibinfo {author} {\bibfnamefont {M.~R.}\ \bibnamefont {Rosenberger}}, \bibinfo {author} {\bibfnamefont {C.~S.}\ \bibnamefont {Hellberg}}, \bibinfo {author} {\bibfnamefont {S.~V.}\ \bibnamefont {Sivaram}}, \bibinfo {author} {\bibfnamefont {K.~M.}\ \bibnamefont {McCreary}}, \bibinfo {author} {\bibfnamefont {I.~I.}\ \bibnamefont {Mazin}},\ and\ \bibinfo {author} {\bibfnamefont {B.~T.}\ \bibnamefont {Jonker}},\ }\bibfield  {title} {\bibinfo {title} {Double indirect interlayer exciton in a mose2/wse2 van der waals heterostructure},\ }\href {https://doi.org/10.1021/ac- snano.8b01369} {\bibfield  {journal} {\bibinfo  {journal} {ACS nano}\ }\textbf {\bibinfo {volume} {12}},\ \bibinfo {pages} {4719} (\bibinfo {year} {2018})}\BibitemShut {NoStop}%
\bibitem [{\citenamefont {Choi}\ \emph {et~al.}(2021)\citenamefont {Choi}, \citenamefont {Florian}, \citenamefont {Steinhoff}, \citenamefont {Erben}, \citenamefont {Tran}, \citenamefont {Kim}, \citenamefont {Sun}, \citenamefont {Quan}, \citenamefont {Claassen}, \citenamefont {Majumder} \emph {et~al.}}]{choi2021twist}%
  \BibitemOpen
  \bibfield  {author} {\bibinfo {author} {\bibfnamefont {J.}~\bibnamefont {Choi}}, \bibinfo {author} {\bibfnamefont {M.}~\bibnamefont {Florian}}, \bibinfo {author} {\bibfnamefont {A.}~\bibnamefont {Steinhoff}}, \bibinfo {author} {\bibfnamefont {D.}~\bibnamefont {Erben}}, \bibinfo {author} {\bibfnamefont {K.}~\bibnamefont {Tran}}, \bibinfo {author} {\bibfnamefont {D.~S.}\ \bibnamefont {Kim}}, \bibinfo {author} {\bibfnamefont {L.}~\bibnamefont {Sun}}, \bibinfo {author} {\bibfnamefont {J.}~\bibnamefont {Quan}}, \bibinfo {author} {\bibfnamefont {R.}~\bibnamefont {Claassen}}, \bibinfo {author} {\bibfnamefont {S.}~\bibnamefont {Majumder}}, \emph {et~al.},\ }\bibfield  {title} {\bibinfo {title} {Twist angle-dependent interlayer exciton lifetimes in van der waals heterostructures},\ }\href {https://doi.org/10.1103/Phys- RevLett.126.047401} {\bibfield  {journal} {\bibinfo  {journal} {Physical Review Letters}\ }\textbf {\bibinfo {volume} {126}},\ \bibinfo {pages} {047401} (\bibinfo {year} {2021})}\BibitemShut {NoStop}%
\bibitem [{\citenamefont {Lagarde}\ \emph {et~al.}(2014)\citenamefont {Lagarde}, \citenamefont {Bouet}, \citenamefont {Marie}, \citenamefont {Zhu}, \citenamefont {Liu}, \citenamefont {Amand}, \citenamefont {Tan},\ and\ \citenamefont {Urbaszek}}]{lagarde2014carrier}%
  \BibitemOpen
  \bibfield  {author} {\bibinfo {author} {\bibfnamefont {D.}~\bibnamefont {Lagarde}}, \bibinfo {author} {\bibfnamefont {L.}~\bibnamefont {Bouet}}, \bibinfo {author} {\bibfnamefont {X.}~\bibnamefont {Marie}}, \bibinfo {author} {\bibfnamefont {C.}~\bibnamefont {Zhu}}, \bibinfo {author} {\bibfnamefont {B.}~\bibnamefont {Liu}}, \bibinfo {author} {\bibfnamefont {T.}~\bibnamefont {Amand}}, \bibinfo {author} {\bibfnamefont {P.}~\bibnamefont {Tan}},\ and\ \bibinfo {author} {\bibfnamefont {B.}~\bibnamefont {Urbaszek}},\ }\bibfield  {title} {\bibinfo {title} {Carrier and polarization dynamics in monolayer mos 2},\ }\href {https://doi.org/10.1103/PhysRevLett.112.047401} {\bibfield  {journal} {\bibinfo  {journal} {Physical review letters}\ }\textbf {\bibinfo {volume} {112}},\ \bibinfo {pages} {047401} (\bibinfo {year} {2014})}\BibitemShut {NoStop}%
\bibitem [{\citenamefont {Korn}\ \emph {et~al.}(2011)\citenamefont {Korn}, \citenamefont {Heydrich}, \citenamefont {Hirmer}, \citenamefont {Schmutzler},\ and\ \citenamefont {Sch{\"u}ller}}]{korn2011low}%
  \BibitemOpen
  \bibfield  {author} {\bibinfo {author} {\bibfnamefont {T.}~\bibnamefont {Korn}}, \bibinfo {author} {\bibfnamefont {S.}~\bibnamefont {Heydrich}}, \bibinfo {author} {\bibfnamefont {M.}~\bibnamefont {Hirmer}}, \bibinfo {author} {\bibfnamefont {J.}~\bibnamefont {Schmutzler}},\ and\ \bibinfo {author} {\bibfnamefont {C.}~\bibnamefont {Sch{\"u}ller}},\ }\bibfield  {title} {\bibinfo {title} {Low-temperature photocarrier dynamics in monolayer mos2},\ }\bibfield  {journal} {\bibinfo  {journal} {Applied Physics Letters}\ }\textbf {\bibinfo {volume} {99}},\ \href {https://doi.org/10.1063/1.3636402} {10.1063/1.3636402} (\bibinfo {year} {2011})\BibitemShut {NoStop}%
\bibitem [{\citenamefont {Shi}\ \emph {et~al.}(2013)\citenamefont {Shi}, \citenamefont {Yan}, \citenamefont {Bertolazzi}, \citenamefont {Brivio}, \citenamefont {Gao}, \citenamefont {Kis}, \citenamefont {Jena}, \citenamefont {Xing},\ and\ \citenamefont {Huang}}]{shi2013exciton}%
  \BibitemOpen
  \bibfield  {author} {\bibinfo {author} {\bibfnamefont {H.}~\bibnamefont {Shi}}, \bibinfo {author} {\bibfnamefont {R.}~\bibnamefont {Yan}}, \bibinfo {author} {\bibfnamefont {S.}~\bibnamefont {Bertolazzi}}, \bibinfo {author} {\bibfnamefont {J.}~\bibnamefont {Brivio}}, \bibinfo {author} {\bibfnamefont {B.}~\bibnamefont {Gao}}, \bibinfo {author} {\bibfnamefont {A.}~\bibnamefont {Kis}}, \bibinfo {author} {\bibfnamefont {D.}~\bibnamefont {Jena}}, \bibinfo {author} {\bibfnamefont {H.~G.}\ \bibnamefont {Xing}},\ and\ \bibinfo {author} {\bibfnamefont {L.}~\bibnamefont {Huang}},\ }\bibfield  {title} {\bibinfo {title} {Exciton dynamics in suspended monolayer and few-layer mos2 2d crystals},\ }\href@noop {} {\bibfield  {journal} {\bibinfo  {journal} {ACS nano}\ }\textbf {\bibinfo {volume} {7}},\ \bibinfo {pages} {1072} (\bibinfo {year} {2013})}\BibitemShut {NoStop}%
\bibitem [{\citenamefont {Peimyoo}\ \emph {et~al.}(2013)\citenamefont {Peimyoo}, \citenamefont {Shang}, \citenamefont {Cong}, \citenamefont {Shen}, \citenamefont {Wu}, \citenamefont {Yeow},\ and\ \citenamefont {Yu}}]{peimyoo2013nonblinking}%
  \BibitemOpen
  \bibfield  {author} {\bibinfo {author} {\bibfnamefont {N.}~\bibnamefont {Peimyoo}}, \bibinfo {author} {\bibfnamefont {J.}~\bibnamefont {Shang}}, \bibinfo {author} {\bibfnamefont {C.}~\bibnamefont {Cong}}, \bibinfo {author} {\bibfnamefont {X.}~\bibnamefont {Shen}}, \bibinfo {author} {\bibfnamefont {X.}~\bibnamefont {Wu}}, \bibinfo {author} {\bibfnamefont {E.~K.}\ \bibnamefont {Yeow}},\ and\ \bibinfo {author} {\bibfnamefont {T.}~\bibnamefont {Yu}},\ }\bibfield  {title} {\bibinfo {title} {Nonblinking, intense two-dimensional light emitter: monolayer ws2 triangles},\ }\href {https://doi.org/10.1021/nn4046002} {\bibfield  {journal} {\bibinfo  {journal} {ACS nano}\ }\textbf {\bibinfo {volume} {7}},\ \bibinfo {pages} {10985} (\bibinfo {year} {2013})}\BibitemShut {NoStop}%
\bibitem [{\citenamefont {Re~Fiorentin}\ \emph {et~al.}(2021)\citenamefont {Re~Fiorentin}, \citenamefont {Risplendi}, \citenamefont {Palummo},\ and\ \citenamefont {Cicero}}]{re2021first}%
  \BibitemOpen
  \bibfield  {author} {\bibinfo {author} {\bibfnamefont {M.}~\bibnamefont {Re~Fiorentin}}, \bibinfo {author} {\bibfnamefont {F.}~\bibnamefont {Risplendi}}, \bibinfo {author} {\bibfnamefont {M.}~\bibnamefont {Palummo}},\ and\ \bibinfo {author} {\bibfnamefont {G.}~\bibnamefont {Cicero}},\ }\bibfield  {title} {\bibinfo {title} {First-principles calculations of exciton radiative lifetimes in monolayer graphitic carbon nitride nanosheets: implications for photocatalysis},\ }\href {https://doi.org/10.1021/ac- sanm.0c03317} {\bibfield  {journal} {\bibinfo  {journal} {ACS Applied Nano Materials}\ }\textbf {\bibinfo {volume} {4}},\ \bibinfo {pages} {1985} (\bibinfo {year} {2021})}\BibitemShut {NoStop}%
\bibitem [{\citenamefont {Volmer}\ \emph {et~al.}(2023)\citenamefont {Volmer}, \citenamefont {Ersfeld}, \citenamefont {Faria~Junior}, \citenamefont {Waldecker}, \citenamefont {Parashar}, \citenamefont {Rathmann}, \citenamefont {Dubey}, \citenamefont {Cojocariu}, \citenamefont {Feyer}, \citenamefont {Watanabe} \emph {et~al.}}]{volmer2023twist}%
  \BibitemOpen
  \bibfield  {author} {\bibinfo {author} {\bibfnamefont {F.}~\bibnamefont {Volmer}}, \bibinfo {author} {\bibfnamefont {M.}~\bibnamefont {Ersfeld}}, \bibinfo {author} {\bibfnamefont {P.~E.}\ \bibnamefont {Faria~Junior}}, \bibinfo {author} {\bibfnamefont {L.}~\bibnamefont {Waldecker}}, \bibinfo {author} {\bibfnamefont {B.}~\bibnamefont {Parashar}}, \bibinfo {author} {\bibfnamefont {L.}~\bibnamefont {Rathmann}}, \bibinfo {author} {\bibfnamefont {S.}~\bibnamefont {Dubey}}, \bibinfo {author} {\bibfnamefont {I.}~\bibnamefont {Cojocariu}}, \bibinfo {author} {\bibfnamefont {V.}~\bibnamefont {Feyer}}, \bibinfo {author} {\bibfnamefont {K.}~\bibnamefont {Watanabe}}, \emph {et~al.},\ }\bibfield  {title} {\bibinfo {title} {Twist angle dependent interlayer transfer of valley polarization from excitons to free charge carriers in wse2/mose2 heterobilayers},\ }\href {https://doi.org/10.1038/s41699-023-00420-1} {\bibfield  {journal} {\bibinfo  {journal} {npj 2D Materials and Applications}\ }\textbf {\bibinfo {volume} {7}},\
  \bibinfo {pages} {58} (\bibinfo {year} {2023})}\BibitemShut {NoStop}%
\bibitem [{\citenamefont {Miller}\ \emph {et~al.}(2017)\citenamefont {Miller}, \citenamefont {Steinhoff}, \citenamefont {Pano}, \citenamefont {Klein}, \citenamefont {Jahnke}, \citenamefont {Holleitner},\ and\ \citenamefont {Wurstbauer}}]{miller2017long}%
  \BibitemOpen
  \bibfield  {author} {\bibinfo {author} {\bibfnamefont {B.}~\bibnamefont {Miller}}, \bibinfo {author} {\bibfnamefont {A.}~\bibnamefont {Steinhoff}}, \bibinfo {author} {\bibfnamefont {B.}~\bibnamefont {Pano}}, \bibinfo {author} {\bibfnamefont {J.}~\bibnamefont {Klein}}, \bibinfo {author} {\bibfnamefont {F.}~\bibnamefont {Jahnke}}, \bibinfo {author} {\bibfnamefont {A.}~\bibnamefont {Holleitner}},\ and\ \bibinfo {author} {\bibfnamefont {U.}~\bibnamefont {Wurstbauer}},\ }\bibfield  {title} {\bibinfo {title} {Long-lived direct and indirect interlayer excitons in van der waals heterostructures},\ }\href {https://doi.org/10.1021/acs.nanolett.7b01304} {\bibfield  {journal} {\bibinfo  {journal} {Nano letters}\ }\textbf {\bibinfo {volume} {17}},\ \bibinfo {pages} {5229} (\bibinfo {year} {2017})}\BibitemShut {NoStop}%
\bibitem [{\citenamefont {Baranowski}\ \emph {et~al.}(2017)\citenamefont {Baranowski}, \citenamefont {Surrente}, \citenamefont {Klopotowski}, \citenamefont {Urban}, \citenamefont {Zhang}, \citenamefont {Maude}, \citenamefont {Wiwatowski}, \citenamefont {Mackowski}, \citenamefont {Kung}, \citenamefont {Dumcenco} \emph {et~al.}}]{baranowski2017probing}%
  \BibitemOpen
  \bibfield  {author} {\bibinfo {author} {\bibfnamefont {M.}~\bibnamefont {Baranowski}}, \bibinfo {author} {\bibfnamefont {A.}~\bibnamefont {Surrente}}, \bibinfo {author} {\bibfnamefont {L.}~\bibnamefont {Klopotowski}}, \bibinfo {author} {\bibfnamefont {J.~M.}\ \bibnamefont {Urban}}, \bibinfo {author} {\bibfnamefont {N.}~\bibnamefont {Zhang}}, \bibinfo {author} {\bibfnamefont {D.~K.}\ \bibnamefont {Maude}}, \bibinfo {author} {\bibfnamefont {K.}~\bibnamefont {Wiwatowski}}, \bibinfo {author} {\bibfnamefont {S.}~\bibnamefont {Mackowski}}, \bibinfo {author} {\bibfnamefont {Y.-C.}\ \bibnamefont {Kung}}, \bibinfo {author} {\bibfnamefont {D.}~\bibnamefont {Dumcenco}}, \emph {et~al.},\ }\bibfield  {title} {\bibinfo {title} {Probing the interlayer exciton physics in a mos2/mose2/mos2 van der waals heterostructure},\ }\href {https://doi.org/10.1021/acs.nanolett.7b03184} {\bibfield  {journal} {\bibinfo  {journal} {Nano letters}\ }\textbf {\bibinfo {volume} {17}},\ \bibinfo {pages} {6360} (\bibinfo {year}
  {2017})}\BibitemShut {NoStop}%
\bibitem [{\citenamefont {Kiemle}\ \emph {et~al.}(2020)\citenamefont {Kiemle}, \citenamefont {Sigger}, \citenamefont {Lorke}, \citenamefont {Miller}, \citenamefont {Watanabe}, \citenamefont {Taniguchi}, \citenamefont {Holleitner},\ and\ \citenamefont {Wurstbauer}}]{kiemle2020control}%
  \BibitemOpen
  \bibfield  {author} {\bibinfo {author} {\bibfnamefont {J.}~\bibnamefont {Kiemle}}, \bibinfo {author} {\bibfnamefont {F.}~\bibnamefont {Sigger}}, \bibinfo {author} {\bibfnamefont {M.}~\bibnamefont {Lorke}}, \bibinfo {author} {\bibfnamefont {B.}~\bibnamefont {Miller}}, \bibinfo {author} {\bibfnamefont {K.}~\bibnamefont {Watanabe}}, \bibinfo {author} {\bibfnamefont {T.}~\bibnamefont {Taniguchi}}, \bibinfo {author} {\bibfnamefont {A.}~\bibnamefont {Holleitner}},\ and\ \bibinfo {author} {\bibfnamefont {U.}~\bibnamefont {Wurstbauer}},\ }\bibfield  {title} {\bibinfo {title} {Control of the orbital character of indirect excitons in mos 2/ws 2 heterobilayers},\ }\href {https://doi.org/10.1103/PhysRevB.101.121404} {\bibfield  {journal} {\bibinfo  {journal} {Physical Review B}\ }\textbf {\bibinfo {volume} {101}},\ \bibinfo {pages} {121404} (\bibinfo {year} {2020})}\BibitemShut {NoStop}%
\bibitem [{\citenamefont {Rivera}\ \emph {et~al.}(2015)\citenamefont {Rivera}, \citenamefont {Schaibley}, \citenamefont {Jones}, \citenamefont {Ross}, \citenamefont {Wu}, \citenamefont {Aivazian}, \citenamefont {Klement}, \citenamefont {Seyler}, \citenamefont {Clark}, \citenamefont {Ghimire} \emph {et~al.}}]{rivera2015observation}%
  \BibitemOpen
  \bibfield  {author} {\bibinfo {author} {\bibfnamefont {P.}~\bibnamefont {Rivera}}, \bibinfo {author} {\bibfnamefont {J.~R.}\ \bibnamefont {Schaibley}}, \bibinfo {author} {\bibfnamefont {A.~M.}\ \bibnamefont {Jones}}, \bibinfo {author} {\bibfnamefont {J.~S.}\ \bibnamefont {Ross}}, \bibinfo {author} {\bibfnamefont {S.}~\bibnamefont {Wu}}, \bibinfo {author} {\bibfnamefont {G.}~\bibnamefont {Aivazian}}, \bibinfo {author} {\bibfnamefont {P.}~\bibnamefont {Klement}}, \bibinfo {author} {\bibfnamefont {K.}~\bibnamefont {Seyler}}, \bibinfo {author} {\bibfnamefont {G.}~\bibnamefont {Clark}}, \bibinfo {author} {\bibfnamefont {N.~J.}\ \bibnamefont {Ghimire}}, \emph {et~al.},\ }\bibfield  {title} {\bibinfo {title} {Observation of long-lived interlayer excitons in monolayer mose2--wse2 heterostructures},\ }\href {https://doi.org/10.1038/ncomms7242} {\bibfield  {journal} {\bibinfo  {journal} {Nature communications}\ }\textbf {\bibinfo {volume} {6}},\ \bibinfo {pages} {6242} (\bibinfo {year} {2015})}\BibitemShut {NoStop}%
\bibitem [{\citenamefont {Rohlfing}\ \emph {et~al.}(1998)\citenamefont {Rohlfing}, \citenamefont {Kr{\"u}ger},\ and\ \citenamefont {Pollmann}}]{rohlfing1998role}%
  \BibitemOpen
  \bibfield  {author} {\bibinfo {author} {\bibfnamefont {M.}~\bibnamefont {Rohlfing}}, \bibinfo {author} {\bibfnamefont {P.}~\bibnamefont {Kr{\"u}ger}},\ and\ \bibinfo {author} {\bibfnamefont {J.}~\bibnamefont {Pollmann}},\ }\bibfield  {title} {\bibinfo {title} {Role of semicore d electrons in quasiparticle band-structure calculations},\ }\href {https://doi.org/10.1103/Phys- RevB.57.6485} {\bibfield  {journal} {\bibinfo  {journal} {Physical Review B}\ }\textbf {\bibinfo {volume} {57}},\ \bibinfo {pages} {6485} (\bibinfo {year} {1998})}\BibitemShut {NoStop}%
\bibitem [{\citenamefont {Yu}\ \emph {et~al.}(2015)\citenamefont {Yu}, \citenamefont {Cui}, \citenamefont {Xu},\ and\ \citenamefont {Yao}}]{yu2015valley}%
  \BibitemOpen
  \bibfield  {author} {\bibinfo {author} {\bibfnamefont {H.}~\bibnamefont {Yu}}, \bibinfo {author} {\bibfnamefont {X.}~\bibnamefont {Cui}}, \bibinfo {author} {\bibfnamefont {X.}~\bibnamefont {Xu}},\ and\ \bibinfo {author} {\bibfnamefont {W.}~\bibnamefont {Yao}},\ }\bibfield  {title} {\bibinfo {title} {Valley excitons in two-dimensional semiconductors},\ }\href {https://doi.org/10.1093/nsr/nwu078} {\bibfield  {journal} {\bibinfo  {journal} {National Science Review}\ }\textbf {\bibinfo {volume} {2}},\ \bibinfo {pages} {57} (\bibinfo {year} {2015})}\BibitemShut {NoStop}%
\bibitem [{\citenamefont {Wilson}\ \emph {et~al.}(2017)\citenamefont {Wilson}, \citenamefont {Nguyen}, \citenamefont {Seyler}, \citenamefont {Rivera}, \citenamefont {Marsden}, \citenamefont {Laker}, \citenamefont {Constantinescu}, \citenamefont {Kandyba}, \citenamefont {Barinov}, \citenamefont {Hine} \emph {et~al.}}]{wilson2017determination}%
  \BibitemOpen
  \bibfield  {author} {\bibinfo {author} {\bibfnamefont {N.~R.}\ \bibnamefont {Wilson}}, \bibinfo {author} {\bibfnamefont {P.~V.}\ \bibnamefont {Nguyen}}, \bibinfo {author} {\bibfnamefont {K.}~\bibnamefont {Seyler}}, \bibinfo {author} {\bibfnamefont {P.}~\bibnamefont {Rivera}}, \bibinfo {author} {\bibfnamefont {A.~J.}\ \bibnamefont {Marsden}}, \bibinfo {author} {\bibfnamefont {Z.~P.}\ \bibnamefont {Laker}}, \bibinfo {author} {\bibfnamefont {G.~C.}\ \bibnamefont {Constantinescu}}, \bibinfo {author} {\bibfnamefont {V.}~\bibnamefont {Kandyba}}, \bibinfo {author} {\bibfnamefont {A.}~\bibnamefont {Barinov}}, \bibinfo {author} {\bibfnamefont {N.~D.}\ \bibnamefont {Hine}}, \emph {et~al.},\ }\bibfield  {title} {\bibinfo {title} {Determination of band offsets, hybridization, and exciton binding in 2d semiconductor heterostructures},\ }\href {https://doi.org/10.1126/sciadv.1601832} {\bibfield  {journal} {\bibinfo  {journal} {Science advances}\ }\textbf {\bibinfo {volume} {3}},\ \bibinfo {pages} {e1601832} (\bibinfo
  {year} {2017})}\BibitemShut {NoStop}%
\end{thebibliography}%
\end{document}